\documentclass[prx,preprint,showpacs,preprintnumbers,amsmath,amssymb,floatfix]{revtex4-1}
\usepackage{graphicx}

\usepackage{bm}
\usepackage{xcolor}
\usepackage{url}
\usepackage{bookmark}

\newcommand{\be}{\begin{equation}}
\newcommand{\ee}{\end{equation}}
\newcommand{\eq}[1]{Eq.~(\ref{#1})}
\newcommand{\fig}[1]{Fig.~\ref{#1}}
\def\bea{\begin{eqnarray}}
\def\eea{\end{eqnarray}}

\def\bra{\langle}
\def\ket{\rangle}
\def\vq{{\bf q}}

\def\vk{{\bf k}}
\def\vQ{{\bf Q}}
\def\vr{{\bf r}}

\begin{document}

\title{Spin and bond-charge excitation spectra in correlated electron systems near an antiferromagnetic phase} 

\author{Muhammad Zafur$^{1,2}$ and Hiroyuki Yamase$^{2.1}$}
\affiliation{
{$^1$}Department of Condensed Matter Physics, Graduate School of Science, Hokkaido University, Sapporo 060-0810, Japan  \\
{$^2$}Research Center for Materials Nanoarchitectonics (MANA), National Institute for Materials Science (NIMS), Tsukuba 305-0047, Japan
}

\date{February 29, 2024}

\begin{abstract}
Magnetic and bond-charge interactions can arise from the same microscopic interaction. Motivated by this observation, we compute magnetic and bond-charge excitation spectra on an equal footing by introducing a simple effective model on a square lattice, which describes antiferromagnetic and $d$-wave superconducting phases around half-filling on the electron-doped side. The magnetic excitation spectrum Im$\chi(\vq, \omega)$ has strong weight around $\vq=(\pi, \pi)$ in low energy and its intensity map exhibits a pencil-tip-like shape in $\vq$-$\omega$ space. Around $\vq=(0,0)$ magnetic excitations show a steep dispersion toward the $(\pi, \pi)$ and $(\pi,0)$ directions, which is very similar to a spin-wave dispersion although the system is non-magnetic. Bond-charge excitations are characterized by four different symmetries and studied for all possible couplings. Bond-charge fluctuations with three different symmetries have large spectral weight around $\vq=(\pi, \pi)$ in a relatively low-energy region and extend widely more than the  magnetic excitation spectrum. The $d$-wave symmetry of bond-charge excitations also has sizable spectral weight along the direction $(\pi/2, \pi/2)$-$(0, 0)$-$(\pi/2, 0)$ in a low-energy region and exhibits softening around $\vq \approx (0.5 \pi, 0)$, whereas no such softening is present in the other symmetries. These results capture the essential features observed in electron-doped cuprates and may motivate an experimental test of bond-charge excitations around $\vq=(\pi, \pi)$ on top of the strong magnetic excitations there as well as additional softening in the $d$-wave channel in the $(\pi, \pi)$-$(\pi/2, \pi/2)$ region at low temperatures near the magnetic phase. We extend the present analysis to the hole-doped side and highlight a contrast to the electron-doped side, which includes incommensurate correlations, electronic nematic correlations, and spin and bond-charge resonance modes in the superconducting state. 
\end{abstract}


\maketitle

\section{introduction}
High-temperature cuprate superconductivity is realized by  carrier doping in antiferromagnetic Mott insulators \cite{keimer15} and the spin degree of freedom is believed to carry crucial information to understand the mechanism of high-temperature superconductivity \cite{anderson87}.  Magnetic excitation spectra in cuprates were extensively studied by inelastic neutron scattering \cite{birgeneau06,fujita12} including a high-energy region \cite{ishii14,wakimoto15}.  Large spectral weight is found around the momentum $\vq=(\pi, \pi)$ in a low-energy region \cite{birgeneau06,fujita12}, much smaller than the Heisenberg spin exchange interaction strength $J$, which is about $100$-$150$ meV in the parent compound of cuprates \cite{bourges97b}. The presence of the low-energy magnetic fluctuations provides a solid basis to discuss the high-$T_{c}$ mechanism in terms of magnetic fluctuations \cite{scalapino12}. Besides the large spectral weight around $\vq=(\pi, \pi)$, recently resonant inelastic x-ray scattering (RIXS) revealed spin-wave-like excitations around $\vq=(0,0)$ up to a rather high-energy region even in the paramagnetic phase \cite{braicovich10,letacon11,letacon13,dean13,ishii14,wslee14,wakimoto15,monney16,ivashko17,meyers17,chaix18,peng18a,robarts19,wang22}.

Needless to say, the understanding of the charge dynamics should be equally important. In fact, it is mobile charged electrons which yield superconductivity. While neutron scattering experiments reported the magnetic excitation spectrum in space of momentum $\vq$ and energy $\omega$ already around 1990 \cite{thurston89,rossat-mignod91}, it was much later when the charge excitation spectrum was revealed by RIXS in $\vq$-$\omega$ space around mainly two different momenta: $\vq \approx (0,0)$ \cite{ishii05,ishii14,wslee14,ishii17,dellea17,hepting18,lin20,nag20,hepting22,singh22a,hepting23} and $(q_{x}, 0)$ \cite{ghiringhelli12,hashimoto14,peng16,chaix17,arpaia19,yu20,wslee21,lu22,arpaia23}.   

While the origin of the charge excitations around $\vq=(0,0)$ was discussed controversially in the early days  \cite{ishii05,ishii14,wslee14,greco16,ishii17, dellea17}, they are now consistently interpreted as acoustic-like plasmons, but with a gap at $\vq=(0,0)$ \cite{greco16,hepting22}. The acoustic-like plasmons were observed in both electron-doped  \cite{hepting18,lin20,hepting22} and hole-doped cuprates \cite{nag20,singh22a,hepting23}. These plasmons have a finite momentum $q_{z}$ along the $z$ direction and the conventional optical plasmon  \cite{nuecker89,romberg90,bozovic90} corresponds to the plasmon branch at $q_{z}=0$. The large-$N$ theory of the layered $t$-$J$ model with the long-range Coulomb interaction explained accurately the observed plasmon excitation spectra in different cuprate materials  by choosing realistic  band parameters \cite{greco19,greco20,nag20,hepting22, hepting23}. 

The charge excitations along the direction $(q_{x},0)$ are characterized by their softening at $q_{x} \sim 0.6\pi (0.5\pi)$ for hole-doped \cite{ghiringhelli12,chang12,achkar12,blackburn13,blanco-canosa14,comin14,tabis14,da-silva-neto14,hashimoto14,peng16,chaix17,arpaia19,yu20,wslee21,lu22,arpaia23,misc-LSCO} (electron-doped \cite{da-silva-neto15,da-silva-neto16,da-silva-neto18}) cuprates. Despite numerous studies in hole-doped cuprates, their origin is still controversial \cite{bejas12,allais14,meier14,wang14,atkinson15,yamakawa15,mishra15,zeyher18}. On the other hand, for electron-doped cuprates, theoretical studies suggested that those charge excitations come from the $d$-wave bond-charge ordering tendency \cite{yamase15b,li17} and explained their doping and temperature dependences observed in experiments \cite{yamase19}.  

Important insight of the charge dynamics in cuprates obtained so far is that there are at least two different types of charge excitations \cite{yamase21c}: one is plasmons and the other is bond-charge fluctuations. To address the former, one needs to take into account not only the long-range Coulomb interaction but also the layered structure of cuprates to describe the singular behavior of the long-range Coulomb interaction correctly  \cite{greco16,greco19,greco20,nag20,hepting22,hepting23}. In contrast, the latter can be captured already in the conventional two-dimensional (2D) model of cuprates \cite{yamase15b,li17,yamase19}. On top of these fundamental differences,  plasmons and bond-charge excitations are characterized by the different energy scale. The bond-charge excitations occur typically inside the particle-hole continuum whereas plasmons are well defined above the continuum. In this sense, they are decoupled to each other and can be addressed separately. In fact, the origin of the bond-charge order lies in the Heisenberg interaction $J$ \cite{fczhang88a, affleck88,marston89,morse91,cappelluti99,yamase00a,yamase00b,lee06,bejas12} or more broadly in an effective spin-spin interaction \cite{sachdev13} whereas that of plasmons lies in the singularity of the long-range Coulomb interaction in the long-wavelength limit \cite{giuliani}. 

In particular, we are interested in bond-charge excitations controlled by the spin interaction, because the same interaction also controls magnetic excitations and needless to say, superconductivity \cite{anderson87,scalapino12}. The magnetic fluctuations are well  known to have strong spectral weight around $\vq=(\pi,\pi)$ at low energy \cite{birgeneau06,fujita12} and a steep dispersive feature around $\vq=(0,0)$ \cite{braicovich10,letacon11,letacon13,dean13,ishii14,wslee14,wakimoto15,monney16,ivashko17,meyers17,chaix18,peng18a,robarts19,wang22}---often referred to as paramagnons (we shall argue that this referring may be misleading in Sec.~IV F). It is then natural to ask: how is the spectral weight of bond-charge fluctuations described in $\vq$-$\omega$ space? Are they similar to the magnetic excitations?

To address these questions, we need to handle both magnetic and bond-charge excitations on an equal footing in the same model. While magnetic excitations were widely studied in the 2D Hubbard-like model \cite{bulut96,norman00,manske01,schnyder04,jia14,arovas22,qin22}, the $t$-$J$ model \cite{tanamoto94,brinckmann99,yamase01,brinckmann02,onufrieva02,li02,sega03,li03,yuan05,yamase06,yamase07,chen10,li16}, and other schemes \cite{si93,zha93,liu95,millis96,kao00,chubukov01,onufrieva04,james12,zhang13}, bond-charge excitations were studied mainly in the  $t$-$J$ model. 

In the slave-boson formalism, a staggered flux phase was introduced and its fluctuations were incorporated in the SU(2) gauge theory \cite{wen96,palee98}.  Strictly speaking, the staggered flux order is defined in terms of spinons and thus is not a bond-charge order. However, when holons are condensed, it is reduced to the so-called $d$-density wave \cite{chakravarty01} or equally flux phase \cite{cappelluti99}. The excitation spectrum of the $d$-density wave was, however, not studied in $\vq$-$\omega$ space. On the other hand, the excitation spectrum of $d$-wave bond-charge order, which becomes the electronic nematic order in the limit of $\vq=(0,0)$ \cite{yamase00a,yamase00b,metzner00}, was studied close to the $d$-wave Pomeranchuk instability in the context of possible competition with $d$-wave superconducting order \cite{yamase04b}. In the slave-boson formalism, the spin degree of freedom is carried by spinons and the charge is by holons. Such a theory contains an emergent U(1) or SU(2) gauge field \cite{lee06}, which yields additional complications. It is not straightforward to handle both spin and charge degrees of freedom on an equal footing. 

On the other hand, in the large-$N$ formalism of the $t$-$J$ model, spin and charge degrees of freedom are carried by the same electrons and furthermore the effect of the local constraint---no double occupancy of electrons at any lattice site---can be incorporated already at leading order. Bond-charge excitations were studied comprehensively in $\vq$-$\omega$ space \cite{bejas17}. Three major channels were studied: $d$-wave and $s$-wave bond-charge, and $d$-density-wave. All of them have strong spectral weight around $\vq=(\pi, \pi)$ in a low-energy region, where magnetic excitations are also expected to have strong spectral weight. In the large-$N$ theory, however, the effect of spin fluctuations is reduced by $O(1/N)$ compared with that of the charge sector. Hence, it is difficult to study both spin and charge degrees of freedom on an equal footing. 

Recently, both magnetic and charge excitations were studied on an equal footing by a variational method \cite{fidrysiak21}. In this study, however, they analyzed the usual on-site charge excitations including plasmons, not bond-charge excitations.  

Since both magnetic and bond-charge excitations are described by the same energy scale, we wish to clarify how their spectral weight is distributed in momentum-energy space and elucidate similarities and contrasts. To do so, we need a model, which allows us to analyze both magnetic and bond-charge interactions within the same approximation scheme. Given that the concept of the bond-charge excitations has not been widely recognized yet, such a theoretical study will be a pedagogical guide to explore bond-charge excitations. 

We first focus on electron-doped cuprates rather than hole-doped ones. There are various reasons. i) The charge-ordering tendency reported in electron-doped cuprates \cite{da-silva-neto15,da-silva-neto16,da-silva-neto18} is interpreted as $d$-wave bond-charge order \cite{yamase15b,li17,yamase19}, whereas that in hole-doped cuprates is still controversial \cite{bejas12,allais14,meier14,wang14,atkinson15,yamakawa15,mishra15,zeyher18}. ii) In the hole-doped cuprates, the presence of the pseudogap is well established and charge orders or their tendencies were reported inside the pseudogap phase \cite{keimer15}. Hence the effect of the pseudogap seems crucially important to the charge dynamics \cite{loret19}, but the microscopic origin of the pseudogap has been a long-standing issue in the cuprate research \cite{timusk99}. In the electron-doped side, on the other hand, the corresponding pseudogap has not been found at least in the doping region where the superconductivity is realized at low temperatures \cite{armitage10}. In this sense, one may reasonably analyze the electron-doped cuprates without considering the pseudogap, making a theoretical study much simpler and more transparent. iii) Compared with the hole-doped side, electron correlation effects seem not to be so strong even though electron correlation effects themselves are believed to be important. This insight may corroborate the possible absence of the pseudogap as mentioned above and a theoretical indication that the magnetism can be described by the Slater insulators, not the Mott insulators \cite{weber10a,weber10b}. Furthermore, a weak coupling theory may capture some features of the spin and charge excitations in the electron-doped cuprates as implied by the analysis given in Ref.~\cite{kyung04}. 

In this paper, we introduce an effective model with three interaction channels: magnetism, bond charge, and superconductivity. By choosing appropriate parameters, this model can capture essential features of the phase diagram in the electron-doped cuprates. Magnetic excitation spectra consist of two major features. One is a pencil-tip shape distribution of the spectral weight centered at $\vq=(\pi, \pi)$ at low energy. The other is a dispersive signal near the edge of the particle-hole continuum, which is very similar to spin-wave-like excitations developing from $\vq=(0,0)$, but could not be associated with paramagnons. For bond-charge channels, we consider all possible symmetries as well as coupling among different symmetries. Typically their spectra have large spectral weight around $\vq=(\pi, \pi)$ in a low energy region. In addition,  the $d$-wave bond-charge excitations exhibit softening at $\vq \approx (0.5\pi, 0)$ along the direction $(0,0)$-$(\pi,0)$.  A comparison of the magnetic and bond-charge excitation spectra shows that the spectral weight of bond-charge excitations is distributed in much wider $\vq$-$\omega$ space than magnetic excitations, implying a dominant role of bond-charge fluctuations. We also extend the present analysis to the hole-doped side and highlight a contrast to the electron-doped side. 

 The present manuscript is organized as follows. In Sec.~II we introduce an effective model describing the antiferromagnetism, $d$-wave superconductivity, and various bond-charge orders close to the antiferromagnetic phase. We then derive formulas of the dynamical magnetic and bond-charge susceptibilities. Numerical results are given in Sec.~III by first focusing on the electron-doped side and then on the hole-doped side. In each case, we present Fermi surfaces, a phase diagram in the plane of the electron density and temperature, and $\vq$-$\omega$ maps of magnetic and bond-charge excitation spectra. Obtained results are discussed in Sec.~IV and our conclusions are given in Sec.~V. In Appendix, we provide spectra of all components of the bond-charge susceptibilities.

\section{Hamiltoninan and formalism} 
\subsection{Hamiltonian}
To study both magnetic and bond-charge excitations on an equal footing, we introduce the following Hamiltonian on a square lattice, 
\be
H = - \sum_{ij\sigma} t_{ij} c_{i \sigma}^{\dagger}  c_{j \sigma} 
+V_b \sum_{i \tau} \hat{\chi}_{i \tau}^{\dagger}  \hat{\chi}_{i \tau} 
+V_{m} \sum_{i \tau} \hat{m}_{i} \hat{m}_{i+ \tau} 
+V_{s} \sum_{i \tau} \hat{\Delta}_{i \tau}^{\dagger}  \hat{\Delta}_{i \tau} \,. 
\label{Hamiltonian}
 \ee
Here $c_{i \sigma}^{\dagger}$ and $c_{i \sigma}$ are the creation and annihilation operators of electrons with spin $\sigma$ at site $i$.  $t_{i j}$ is the hopping integral between the first, second, and third nearest neighbor sites, and has a value of $t$, $t'$, and $t''$, respectively---otherwise $t_{i j}$ is zero. The second term in \eq{Hamiltonian} describes a bond-charge interaction with the interaction strength $V_{b}$ and $\hat{\chi}_{i \tau}$ is the bond-charge operator defined as 
\be
\hat{\chi}_{i \tau} = \sum_{\sigma} c_{i \sigma}^{\dagger}  c_{i+\tau \sigma} 
\label{def-bond}
\ee
with $\tau=x$ and $y$. Physically this operator denotes the hopping between the nearest-neighbor sites in the $x$ or $y$ direction. The third term is the spin interaction between the nearest-neighbor sites and the magnetization is given by 
\be
\hat{m}_{i}= \frac{1}{2} \sum_{\sigma}\sigma c_{i \sigma}^{\dagger}  c_{i \sigma}  \,.
\ee
When $V_{m}$ is chosen as a positive value, it favors the antiferromagnetic order with momentum $\vQ=(\pi, \pi)$. The last term corresponds to the Cooper interaction in the singlet channel and 
\be
\hat{\Delta}_{i \tau} = c_{i \uparrow} c_{i+\tau \downarrow} - c_{i \downarrow} c_{i+\tau \uparrow} \,.
\ee

In \eq{Hamiltonian} we neglect the transverse spin interaction, but this is not relevant to our study as long as we focus on the non-magnetic phase where the longitudinal magnetic excitations are the same as the transverse ones. 

A choice of $V_{b}$, $V_{m}$, and $V_{s}$ is arbitrary in the condition of $V_{b} \leq 0 $, $V_{m} \geq 0$, and $V_{s} \leq 0$ to ensure that the system is close to the bond-charge, magnetic, and superconducting instabilities. It is interesting to point out that the interaction terms in \eq{Hamiltonian} are formally the same as ones obtained in the slave-boson formalism in the $t$-$J$ model when we choose $V_{b}= -3J/8$, $V_{m}=J$, and $V_{s}=-3J/8$ \cite{lee06}---note that both magnetic and bond-charge degrees of freedom have the same energy scale, which is the property of the $J$ term \cite{fczhang88a, affleck88,marston89,morse91,cappelluti99,yamase00a,yamase00b,lee06,bejas12}, not due to some approximation. In the present model, however, there is no constraint on the Hilbert space and thus the strong correlation effect---no double occupancy of electrons at any lattice site---is not included, in contrast to the case of the $t$-$J$ model.

We analyze \eq{Hamiltonian} by  introducing the following mean fields: 
\bea
&&\chi_{\tau} = \bra  \hat{\chi}_{i \tau}  \ket \,, 
\label{MF-chi}\\
&&m=\bra \hat{m}_{i} \ket {\rm e}^{i \vQ \cdot \vr_{i}}  \,,  \\
&&\Delta_{\tau} = \bra \hat{\Delta}_{i \tau} \ket \,,
\eea
in the bond-charge channel, the magnetic channel, and the pairing channel, respectively.  These mean fields are assumed to be real and uniform, independent of sites $i$. The magnetic phase is characterized by the momentum $\vQ$, namely the standard collinear phase. We can check that the stable solution is $\chi_{x}=\chi_{y} =\chi_{b}$ and $\Delta_{x}= -\Delta_{y} = \Delta_{0}$. After Fourier transformation, the mean-field Hamiltonian is obtained as 
\be
H_{\rm MF} =  \sideset{}{'}\sum_{\vk} \Psi_{\vk}^{\dagger} M_{\vk} \Psi_{\vk} + {\rm const.} \,,
\label{HMF}
\ee
where the $\vk$ summation is restricted to the magnetic Brillouin zone $| k_{x}| + | k_{y}| \leq \pi$ as indicated by a prime, 
\be
\Psi_{\vk}^{\dagger} = ( c_{\vk \uparrow}^{\dagger} , c_{-\vk \downarrow}, c_{\vk+\vQ \uparrow}^{\dagger} , c_{-\vk-\vQ \downarrow}) \,,
\ee 
\be
M_{\vk}=
 \left(
 \begin{array}{cccc}
   \xi_{\bf k} & -\Delta_{\bf k}& -\overline{m} & 0 \\
-\Delta_{\bf k} & -\xi_{\bf k} & 0  & -\overline{m} \\
-\overline{m} & 0 & \xi_{{\bf k}+{\bf Q}} & -\Delta_{{\bf k}+{\bf Q}}\\
0 & -\overline{m}  &  -\Delta_{{\bf k}+{\bf Q}}& -\xi_{{\bf k}+{\bf Q}} 
\end{array}
\right) \,,
\label{Mk}
\ee
and 
\be
{\rm const.} = -N (\mu + 2V_{s}\Delta_{0}^{2}  + 2V_{b} \chi_{b}^{2} - 2V_{m}m^{2})\,.
\ee
Here $N$ is the total number of the lattice sites, 
\bea
&&\xi_{\vk} = -2 \left[ 
(t-V_{b} \chi_{b}) (\cos k_{x} + \cos k_{y}) + 
2t' \cos k_{x} \cos k_{y} + 
t'' (\cos 2k_{x} + \cos 2 k_{y)}
\right] -\mu \,,\\
&&\Delta_{\vk} = 2 V_{s} \Delta_{0} d_{\vk} \,, \\
&& \overline{m} = 2V_{m} m \,, \\
&& d_{\vk} = \cos k_{x} - \cos k_{y}\,,
\eea
and we have introduced the chemical potential term $-\mu \sum_{\vk \sigma} c_{\vk \sigma}^{\dagger} c_{\vk \sigma}$ in \eq{Hamiltonian}. For a later convenience, we also introduce a notation 
\bea
&&s_{\vk} = \cos k_{x}  + \cos k_{y} \,,   \label{formfactor-s} \\
&&p_{\vk}^{\pm} = \sin k_{x}  \pm \sin k_{y} \,.
\label{formfactor}
\eea

It is straightforward to compute the grand canonical potential per lattice site from the mean-field Hamiltonian \eq{HMF}, 
\be
\omega = - \frac{2T} {N} \sideset{}{'}\sum_{\vk} \left[
\log \left(  2 \cosh \frac{\lambda_{\vk}^{+}}{2 T} \right) + 
\log \left(  2 \cosh \frac{\lambda_{\vk}^{-}}{2 T} \right) \right] 
-\mu -2 ( V_{b} \chi_{b}^{2}  + V_{s}\Delta_{0}^{2}  - V_{m}m^{2}) \,.
\label{freeenergy}
\ee
Here $\lambda_{\vk}^{\pm}$ are eigenvalues of the matrix \eq{Mk} and  is given by 
\be
\lambda_{\vk}^{\pm}= \sqrt{ \left(\eta_{\vk}^{\pm}\right)^{2} + \Delta_{\vk}^{2}} \,,
\ee
where 
\bea
&&\eta^{\pm}_{\vk} = \xi^{+}_{\vk} \pm D_{\vk} \,, \\
&&D_{\vk}= \sqrt{ (\xi^{-}_{\vk})^{2} + \overline{m}^{2} } \,, \\
&&\xi^{\pm}_{\vk} = \frac{1}{2} (\xi_{\vk} \pm \xi_{\vk+\vQ} ) \,.
\eea
The functional form $\lambda^{\pm}_{\vk}$ describes that the quasiparticle dispersion $\eta^{\pm}_{\vk}$ in the magnetic phase acquires the pairing gap $\Delta_{\vk}$. 

By minimizing the grand canonical potential \eq{freeenergy} with respect to $\chi_{b}$, $\Delta_{0}$, and $m$, we obtain the following self-consistency equations: 
\bea
&&\chi_{b}=-\frac{1}{2N} \sideset{}{'}\sum_{\vk} \frac{s_{\vk} \xi_{\vk}^{-}}{D_{\vk}} 
\left( 
\frac{\eta^{+}_{\vk}}{\lambda_{\vk}^{+}} \tanh \frac{\lambda_{\vk}^{+}}{2T} - 
\frac{\eta^{-}_{\vk}}{\lambda_{\vk}^{-}} \tanh \frac{\lambda_{\vk}^{-}}{2T} 
\right) \,,\\
&&\Delta_{0}=-\frac{1}{2N} \sideset{}{'}\sum_{\vk} d_{\vk} 
\left( f
\frac{\Delta_{\vk}}{\lambda_{\vk}^{+}} \tanh \frac{\lambda_{\vk}^{+}}{2T} + 
\frac{\Delta_{\vk}}{\lambda_{\vk}^{-}}  \tanh \frac{\lambda_{\vk}^{-}}{2T} 
\right) \,,\\
&& m=\frac{1}{2N} \sideset{}{'}\sum_{\vk} \frac{\overline{m}}{D_{\vk}} 
\left( 
\frac{\eta^{+}_{\vk}}{\lambda_{\vk}^{+}} \tanh \frac{\lambda_{\vk}^{+}}{2T} - 
\frac{\eta^{-}_{\vk}}{\lambda_{\vk}^{-}} \tanh \frac{\lambda_{\vk}^{-}}{2T} 
\right) \,.
\eea
The electron density $n=-\partial{\omega} / \partial{\mu}$ is computed as 
\be
n=1-\frac{1}{N} \sideset{}{'}\sum_{\vk} 
\left( 
\frac{\eta^{+}_{\vk}}{\lambda_{\vk}^{+}} \tanh \frac{\lambda_{\vk}^{+}}{2T} +  
\frac{\eta^{-}_{\vk}}{\lambda_{\vk}^{-}} \tanh \frac{\lambda_{\vk}^{-}}{2T} 
\right) \,.
\ee

\subsection{Dynamical magnetic and bond-charge susceptibilities} 
\subsubsection{General consideration} 
We are interested in studying both magnetic and bond-charge excitations on an equal footing in the same approximation scheme. Here we derive their dynamical susceptibilities in the random phase approximation (RPA) in a non-magnetic state by employing the effective model \eq{Hamiltonian}. Calculations in a magnetic phase would not be performed  appropriately, because the transverse magnetic interaction is neglected in our model. In addition, technically more involved calculations would be required in the magnetic phase \cite{kuboki17,yamase21b}, which is beyond the scope of the present work. 

The presence of the Cooper channel in Hamiltonian \eq{Hamiltonian} leads to the superconducting instability at low temperatures. Hence we rewrite the interaction terms in terms of the Nambu basis $\Phi_{\vk}^{\dagger} = (c^{\dagger}_{\vk \uparrow}, c_{-\vk \downarrow})$. 

Focusing on the bond-charge interaction in \eq{Hamiltonian} and performing the Fourier transformation, we obtain 
\be
H_{b}= \frac{V_{b}}{N} \sum_{\vk, \vk', \vq}\sum_{\sigma, \sigma'} 
(\cos (k_{x} -k_{x}^{'}) + \cos (k_{y} -k_{y}^{'})) 
 c_{\vk  - \frac{\vq}{2} \sigma}^{\dagger}  c_{\vk  + \frac{\vq}{2} \sigma}
 c_{\vk'  + \frac{\vq}{2} \sigma'}^{\dagger}  c_{\vk' - \frac{\vq}{2} \sigma'} \,.
\label{Hbond}
\ee
The form factor can be written in a separable form: 
\be
\cos (k_{x} -k_{x}^{'}) + \cos (k_{y} -k_{y}^{'}) = \frac{1}{2} ( s_{\vk} s_{\vk'} + d_{\vk} d_{\vk'} + 
p_{\vk}^{+} p_{\vk'}^{+} + p_{\vk}^{-} p_{\vk'}^{-}) \, .
\ee
By considering the spin indices, we obtain 
\be
H_{b}= \frac{V_{b}}{2N} \sum_{\vk, \vk', \vq} \sum_{i=1}^{4} 
\left(
\Phi^{\dagger}_{\vk - \frac{\vq}{2}} \hat{\Gamma}_{i}(\vk) 
 \Phi_{\vk + \frac{\vq}{2}} 
\right) 
\left(
\Phi^{\dagger}_{\vk' + \frac{\vq}{2}} 
\hat{\Gamma}_{i}(\vk')   \Phi_{\vk' - \frac{\vq}{2}} 
\right) \, ,
\label{Hbond2}
\ee
where $\hat{\Gamma}_{1}(\vk)=s_{\vk} \sigma_{3}$,  $\hat{\Gamma}_{2}(\vk)=d_{\vk} \sigma_{3}$, $\hat{\Gamma}_{3}(\vk)=p^{+}_{\vk} \sigma_{0}$, $\hat{\Gamma}_{4}(\vk)=p^{-}_{\vk} \sigma_{0}$, and the Pauli matrices, $ \sigma_{3}=
\begin{pmatrix} 
1 & 0 \\ 0&-1 
\end{pmatrix}
$ 
and $ \sigma_{0}=
\begin{pmatrix} 
1 & 0 \\ 0&1 
\end{pmatrix}
$. The difference of the matrices is due to the fact that $s_{\vk}$ and $d_{\vk}$ are even functions with respect to $\vk \leftrightarrow -\vk$ whereas $p^{\pm}_{\vk}$ is an odd function. 
The bond-charge order with $\gamma_{\vk}=p_{\vk}^{-}$ is often called as flux phase \cite{cappelluti99} or $d$-density wave \cite{chakravarty01}, instead of $p_{\vk}^{-}$-wave bond-charge. A possible reason may be the following. Its ordering vector is usually at $\vq=(\pi, \pi)$. In this case, we then shift the variable $\vk$ by $\vq /2$ in \eq{formfactor} and obtain $p_{\vk+\vq/2}^{-}$, which is now $\cos k_{x} - \cos k_{y}$---hence the $d$-wave-like density wave. 

Similarly the magnetic interaction can be written as 
\be
H_{m}= \frac{1}{2N} \sum_{\vk, \vk', \vq} V_{m}(\vq)
\left(
\Phi^{\dagger}_{\vk - \frac{\vq}{2}} \frac{\sigma_{0}}{2}  \Phi_{\vk + \frac{\vq}{2}} 
\right) 
\left(
\Phi^{\dagger}_{\vk' + \frac{\vq}{2}}  \frac{\sigma_{0}}{2}  \Phi_{\vk' - \frac{\vq}{2}} 
\right) \, ,
\ee 
where 
\be
V_{m}(\vq) = 2 V_{m} s_{\vq} \,.
\ee

\begin{figure}[th]
\centering 
\includegraphics[width=10cm]{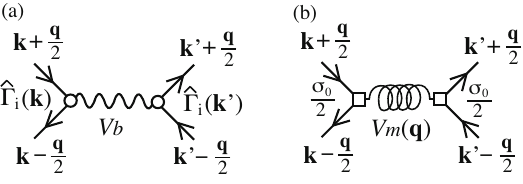}
\caption{Bond-charge (a) and magnetic (b) interaction in the Nambu formalism.  The vertex is given by $\hat{\Gamma}_{i}(\vk)$ in (a) and $\frac{\sigma_{0}}{2}$ in (b) with the interaction strength of $V_{b}$ and $V_{m}(\vq)$, respectively. The solid line denotes the 2 $\times$ 2 Green's function given in \eq{Green}.}
\label{diagram1}
\end{figure}

We do not study fluctuations from the pairing interaction in this work. Instead we include the effect of the pairing interaction by considering a possible presence of the superconducting order. Hence the interaction terms we shall consider are described diagramatically in \fig{diagram1} and the electron Green's function becomes a $ 2 \times 2$ matrix defined in the Nambu space, namely $\hat{\mathcal{G}}_{0}(\vk, \tau) = - \bra  T_{\tau} \Phi_{\vk}(\tau) \Phi_{\vk}^{\dagger}(0) \ket$. After the Fourier transformation, we obtain 
\be
\hat{\mathcal{G}}_{0}(\vk, i k_{n}) = 
\left(
\begin{array}{cc}
\frac{\frac{1}{2}\left( 1+ \frac{\xi_{\vk}}{E_{\vk}} \right)} {i k_{n} - E_{\vk}} +
\frac{\frac{1}{2}\left( 1- \frac{\xi_{\vk}}{E_{\vk}} \right)} {i k_{n} + E_{\vk}} & 
-\frac{\Delta_{\vk}}{2 E_{\vk}} \left(  \frac{1}{ik_{n} - E_{\vk}}  - \frac{1}{ik_{n} + E_{\vk}} \right) \\
-\frac{\Delta_{\vk}}{2 E_{\vk}} \left(  \frac{1}{ik_{n} - E_{\vk}}  - \frac{1}{ik_{n} + E_{\vk}} \right) &
\frac{\frac{1}{2}\left( 1- \frac{\xi_{\vk}}{E_{\vk}} \right)} {i k_{n} - E_{\vk}} +
\frac{\frac{1}{2}\left( 1+ \frac{\xi_{\vk}}{E_{\vk}} \right)} {i k_{n} + E_{\vk}} 
\end{array}
\right) \,, 
\label{Green}
\ee
where $E_{\vk}= \sqrt{\xi_{\vk}^{2} + \Delta_{\vk}^{2}}$.

\begin{figure}[bt]
\centering 
\includegraphics[width=12cm]{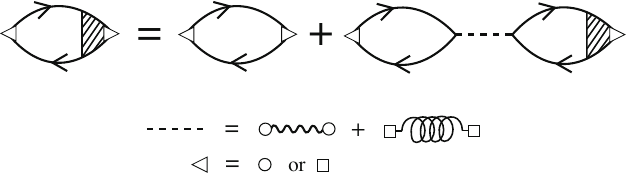}
\caption{Susceptibility in the RPA in the Nambu formalism. There are two channels: bond-charge (wavy line) and magnetic (spring) interactions, although they turn out to be decoupled in the present model; see \fig{diagram3}(a). The open triangle indicates either the open circle or open box---the former describes the bond-charge susceptibility and the latter the magnetic susceptibility. The solid curves denote the Nambu Green's function [\eq{Green}]. }
\label{diagram2}
\end{figure}

\begin{figure}[ht]
\centering
\includegraphics[width=8cm]{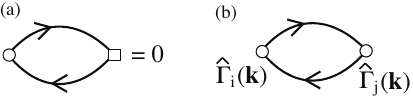}
\caption{(a) Diagram describing a coupling between the bond-charge and magnetic interaction. This diagram vanishes, indicating that these interactions are decoupled in the RPA in \fig{diagram2}. (b) Bubble diagram in the bond-charge channel. The vertices $\hat\Gamma_{i}(\vk)$ and $\hat\Gamma_{j}(\vk)$ can have different symmetries $i \neq j$ in general.} 
\label{diagram3}
\end{figure}

The RPA susceptibility is given by diagrams shown in \fig{diagram2}. These diagrams suggest a possible coupling between the bond-charge and magnetic fluctuations. However, one can show explicitly that a diagram shown in \fig{diagram3}(a) vanishes, indicating that bond-charge and magnetic fluctuations do not couple to each other in the RPA. Intuitively this may be easily understood if one goes back from  the Nambu basis to the usual basis, considers the Green function $\mathcal{G}(\vk, \tau)= - \bra T_{\tau} c_{\vk \sigma}(\tau) c_{\vk \sigma}^{\dagger}(0) \ket$, which does not depend on the spin index, and takes the spin summation in a bubble diagram with vertices, yielding zero---one from the bond-charge channel is even with respect to $\sigma \leftrightarrow -\sigma$ and the other from the magnetic channel is odd. On the other hand, a diagram shown in \fig{diagram3}(b) becomes finite in general. This means that four different symmetries of the vertex $\hat\Gamma_{i}(\vk)$ with $i=1, 2, 3, 4$ can couple to each other---the bond-charge susceptibilities form a $4 \times 4$ matrix.

\subsubsection{Magnetic channel}
The dynamical magnetic susceptibility is obtained by focusing on  the magnetic interaction in \fig{diagram2}, yielding 
\be
\chi(\vq, iq_{m}) = \frac{\chi_{0}(\vq, iq_{m})}{1+ V_{m}(\vq) \chi_{0}(\vq, iq_{m})}\,,
\label{magneRPA}
\ee
where with a notation $\vk_{\pm}=\vk \pm \vq/2$
\bea
&&\chi_{0}(\vq, iq_{m}) = \frac{1}{4N} \sum_{\vk} \left[ 
C^{+}_{\vk, \vq} \left( \tanh \frac{E_{\vk_{-}}}{2T} - \tanh \frac{E_{\vk_{+}}}{2T} \right) 
\frac{1}{E_{\vk_{-}} +iq_{m} - E_{\vk_{+}}} \right.  \nonumber \\
&&\hspace{5mm} 
\left.
+\frac{1}{2}C^{-}_{\vk, \vq} \left( \tanh \frac{E_{\vk_{-}}}{2T} + \tanh \frac{E_{\vk_{+}}}{2T} \right) 
\left( \frac{1}{E_{\vk_{-}} +iq_{m} + E_{\vk_{+}}} + \frac{1}{E_{\vk_{-}} - iq_{m} + E_{\vk_{+}}}
\right) \right]
\label{chi0}
\eea
and the coherent factors are 
\be
C^{\pm}_{\vk, \vq}  = \frac{1}{2} \left( 1 \pm 
\frac{\xi_{\vk_{-}}  \xi_{\vk_{+}} +\Delta_{\vk_{-}} \Delta_{\vk_{+}}} {E_{\vk_{-}} E_{\vk_{+}}}
\right)  \,.
\label{C-factor}  
\ee

\subsubsection{Bond-charge channel}
Since four different symmetries couple to each other, we introduce a $4 \times 4$ bond-charge susceptibility $\hat{\kappa}(\vq, i q_{m})$. The interaction is defined between the same symmetry in \eq{Hbond2}, 
leading to a diagonal matrix: 
\be 
\hat{V}_{b}=
\left(
\begin{array}{cccc}
V_{b} & 0 & 0 & 0 \\
0 & V_{b} & 0 & 0 \\
0 & 0 &  V_{b} &0  \\
0 & 0 & 0 & V_{b}
\end{array}
\right) \,.
\ee
The diagram shown in \fig{diagram2} then yields
\be
\hat{\kappa}(\vq, iq_{m}) = \left(1+ V_{b} \hat{\kappa}_{0}(\vq, iq_{m}) \right)^{-1} \hat{\kappa}_{0}(\vq, iq_{m}) \,,
\label{kappaRPA}
\ee
and each component of $\hat{\kappa}_{0}(\vq, iq_{m})$ is given by 
\be
\hat{\kappa}_{0 \, i j} (\vq, iq_{m}) = - \frac{T}{N} \sum_{\vk, n} T_{r} \hat{\Gamma}_{i}(\vk) \hat{\mathcal{G}}_{0}\left(\vk+\frac{\vq}{2}\right)  \hat{\Gamma}_{j}(\vk) \hat{\mathcal{G}}_{0}\left(\vk - \frac{\vq}{2}\right) \,.
\ee
Here $T_{r}$ is defined for the Nambu space. Since $\hat{\Gamma}_{i}$ is proportional to $\sigma_{0}$ or $\sigma_{3}$, the bond-charge susceptibilities are classified into three categories. 

The first category is for $i,j=1,2$:
\bea
&&\hat{\kappa}_{0 \, i j}(\vq, iq_{m})= \frac{1}{2N} \sum_{\vk} \gamma_{i}(\vk)  \gamma_{j}(\vk) \left[ 
D_{\vk, \vq}^{+} \left(\tanh \frac{E_{\vk_{-}}}{2T} - \tanh \frac{E_{\vk_{+}}}{2T}\right) \right.
\nonumber \\  
&& \hspace{60mm} \left. \times \left( 
\frac{1}{E_{\vk_{-}}+iq_m - E_{\vk_{+}}} + \frac{1}{E_{\vk_{-}} - iq_m - E_{\vk_{+}}} 
\right)  \right. \nonumber \\
&&\hspace{15mm} \left. +  D_{\vk, \vq}^{-} 
\left(\tanh \frac{E_{\vk_{-}}}{2T} + \tanh \frac{E_{\vk_{+}}}{2T}\right)  
\left( 
\frac{1}{E_{\vk_{-}}+iq_m + E_{\vk_{+}}} + \frac{1}{E_{\vk_{-}} - iq_m + E_{\vk_{+}}} 
\right) 
\right]  \,,
\label{kqw1122}
\eea
where $\gamma_{1}=s_{\vk}$ and $\gamma_{2}=d_{\vk}$, which originate from $\hat{\Gamma}_{i}(\vk)$ defined below \eq{Hbond2}, and the coherence factors are 
\be
D_{\vk, \vq}^{\pm} = \frac{1}{2} \left(  1 \pm \frac{\xi_{\vk_{-}} \xi_{\vk_{+}} - \Delta_{\vk_{-}} \Delta_{\vk_{+}}}{ E_{\vk_{-}}  E_{\vk_{+}}} 
\right) \,.
\ee
Note the minus sign in front of $\Delta_{\vk_{-}} \Delta_{\vk_{+}}$, which is different from that in $C_{\vk, \vq}^{\pm}$ in the magnetic channel \eq{C-factor}. 

The second category is for $i,j=3,4$. In this case, we obtain 
\bea
&&\hat{\kappa}_{0 \, i j}(\vq, iq_{m}) = \frac{1}{2N} \sum_{\vk} \bar{\gamma}_{i}(\vk)  \bar{\gamma}_{j}(\vk) \left[ 
C_{\vk, \vq}^{+} \left(\tanh \frac{E_{\vk_{-}}}{2T} - \tanh \frac{E_{\vk_{+}}}{2T}\right) \right.
\nonumber \\  
&& \hspace{60mm} \left. \times \left( 
\frac{1}{E_{\vk_{-}}+iq_m - E_{\vk_{+}}} + \frac{1}{E_{\vk_{-}} - iq_m - E_{\vk_{+}}} 
\right)  \right. \nonumber \\
&&\hspace{15mm} \left. +  C_{\vk, \vq}^{-} 
\left(\tanh \frac{E_{\vk_{-}}}{2T} + \tanh \frac{E_{\vk_{+}}}{2T}\right)  
\left( 
\frac{1}{E_{\vk_{-}}+iq_m + E_{\vk_{+}}} + \frac{1}{E_{\vk_{-}} - iq_m + E_{\vk_{+}}} 
\right) 
\right]  \,.
\label{kqw3344}
\eea
Here $\bar{\gamma}_{3}=p^{+}_{\vk}$, $\bar{\gamma}_{4}=p^{-}_{\vk}$, and the coherence factors are given by \eq{C-factor}. This expression is the same as \eq{chi0} except for the vertex part and the overall factor. 

The third category is for $i=1,2$ and $j=3,4$ or vice versa:
\bea
&&\hat{\kappa}_{0 \, i j}(\vq, iq_{m}) = \frac{1}{2N} \sum_{\vk} \gamma_{i}(\vk)  \bar{\gamma}_{j}(\vk) \left[ 
F_{\vk, \vq}^{+} \left(\tanh \frac{E_{\vk_{-}}}{2T} - \tanh \frac{E_{\vk_{+}}}{2T}\right) \right.
\nonumber \\  
&& \hspace{60mm} \left. \times \left( 
\frac{1}{E_{\vk_{-}}+iq_m - E_{\vk_{+}}} - \frac{1}{E_{\vk_{-}} - iq_m - E_{\vk_{+}}} 
\right)  \right. \nonumber \\
&&\hspace{15mm} \left. +  F_{\vk, \vq}^{-} 
\left(\tanh \frac{E_{\vk_{-}}}{2T} + \tanh \frac{E_{\vk_{+}}}{2T}\right)  
\left( 
\frac{1}{E_{\vk_{-}}+iq_m + E_{\vk_{+}}} - \frac{1}{E_{\vk_{-}} - iq_m + E_{\vk_{+}}} 
\right) 
\right]  \,, 
\eea
where 
\be
F_{\vk, \vq}^{\pm} = \frac{1}{2} \left( \frac{\xi_{\vk_{-}}}{E_{\vk_{-}}}  \pm 
\frac{\xi_{\vk_{+}}}{E_{\vk_{+}}}  \right) \,.
\ee

\section{Results} 
We use the magnetic interaction strength $V_{m}$ as the unit of energy and choose the following parameters: $t/V_{m}=0.7$, $t'/t=-0.14$, $t''/t=0.07$, $V_{b}/V_{m}=-3/8$, and $V_{s}/V_{m}=-3/8$. We first focus on the electron-doped side and have chosen $t'$ and $t''$ to reproduce the Fermi surface observed in electron-doped cuprates, especially NCCO (see the inset in \fig{phase}) \cite{armitage10}. The choice of $V_{b}$ and $V_{s}$ are motivated by the slave-boson formalism of the $t$-$J$ model \cite{lee06}, where the strength of  $V_{b}$ and $V_{s}$ is determined uniquely, although our model does not contain strong correlation effects. The value of $t$ is tuned to reproduce a phase diagram similar to that in electron-doped cuprates \cite{armitage10,fujita03a}. The dynamical magnetic and bond-charge susceptibilities are obtained via the analytical continuation, $iq_{m} \rightarrow \omega + i \Gamma$ in \eq{magneRPA} and \eq{kappaRPA}, respectively. While $\Gamma$ is an infinitesimal with a positive sign, we take $\Gamma=0.1 V_{m}$ for numerical convenience. Phenomenologically $\Gamma$ may simulate a broadening effect from instrumental resolutions and, perhaps, also a partial effect beyond the RPA.

\subsection{Phase diagram}

\begin{figure}[ht]
\centering 
\includegraphics[width=8cm]{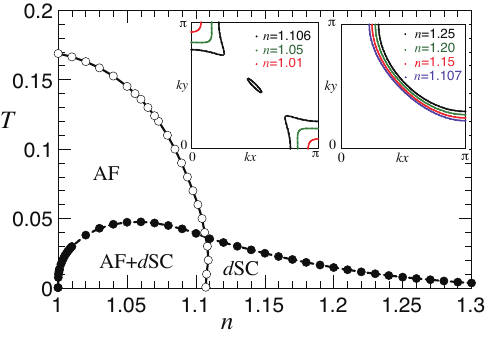}
\caption{Phase diagram in the plane of the electron density $n$ and temperature $T$. The phase boundary is assumed to be continuous. ``$d$SC'' denotes the $d$-wave superconducting phase, ``AF'' the antiferromagnetic phase, and ``AF+$d$SC'' the coexistence of the antiferromagnetism and $d$-wave superconductivity. The energy unit is $V_{m}$. Two insets are Fermi surfaces obtained by assuming $\Delta_{0}=0$ at $T=0.01$ for several choices of the electron density. The Fermi surface is reconstructed in the AF phase (below $n \leq 1.106$) and the electron pockets are realized around $\vk=(\pi,0)$ and $(0,\pi)$; the hole pocket can also be realized around $\vk=(\pi/2, \pi/2)$ close to the phase boundary at $n=1.106$.}
\label{phase}
\end{figure}

Figure~\ref{phase} shows the obtained phase diagram in the plane of the electron density and temperature $T$. Four different phases are stabilized: normal, $d$-wave superconducting, antiferromagnetic, and their coexistence phases. Their phase boundaries are assumed to be continuous. At half-filling ($n=1$), the antiferromagnetic phase is  characterized by the momentum $\vq=(\pi, \pi)$. Moreover, the system is insulating at $n=1$. With electron doping, the system immediately becomes metallic and small pockets are realized around $\vk=(\pi, 0)$ and $(0, \pi)$. Consequently, the coexistence of the $d$-wave superconductivity is stabilized at low temperatures. It is interesting that the maximal value of $T_{c}$ is obtained inside  the antiferromagnetic phase. With further doping, the antiferromagnetic phase disappears, a large Fermi surface is realized, and a pure $d$-wave superconducting state is stabilized at low temperatures. While our Hamiltonian \eq{Hamiltonian} contains the bond-charge interaction, we did not obtain any instability in this channel by confirming that the static $4 \times 4$ bond-charge susceptibilities $\hat{\kappa}(\vq, 0)$ do not diverge  in the parameter region that we are interested in (see Figs.~\ref{kq-T} and \ref{kq-n}).

\begin{figure}[ht]
\centering 
\includegraphics[width=8cm]{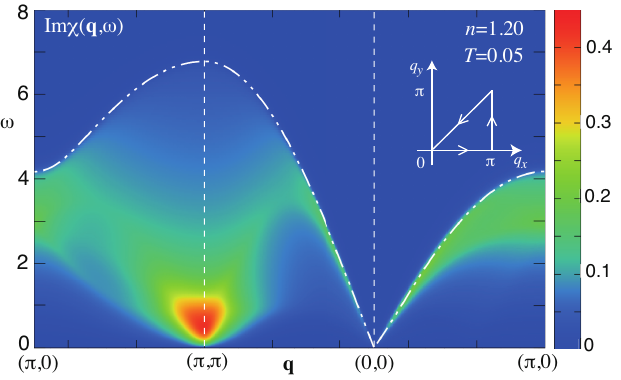}
\caption{$\vq$-$\omega$ map of the spin excitation spectrum Im$\chi(\vq, \omega)$ along the symmetry axis (see the inset) at $n=1.20$ in the normal state ($T=0.05$). The dashed double-dotted curve is the upper boundary of the continuum spectrum.}
\label{xqw-mapT005}
\end{figure}

\subsection{Spin excitation spectrum}

Figure~\ref{xqw-mapT005} is an intensity map of Im$\chi(\vq, \omega)$ in $\vq$-$\omega$ space along the direction $(\pi, 0)$-$(\pi, \pi)$-$(0,0)$-$(\pi, 0)$ in the normal phase. The upper boundary of the continuum spectrum is denoted by a dashed double-dotted curve; the spectrum can extend slightly beyond the boundary because of the broadening $\Gamma$ and a finite temperature effect. 
Inside the continuum,  there is strong spectral weight in low energy around $\vq=(\pi, \pi)$---an indication that the systems is close to the antiferromagnetic phase characterized by $(\pi, \pi)$. There is also a characteristic V-shape dispersion around $\vq=(0,0)$ and relatively strong weight extends up to high energy around $\omega \approx 4$ near $\vq=(\pi,0)$.

\begin{figure}[ht]
\centering 
\includegraphics[width=14cm]{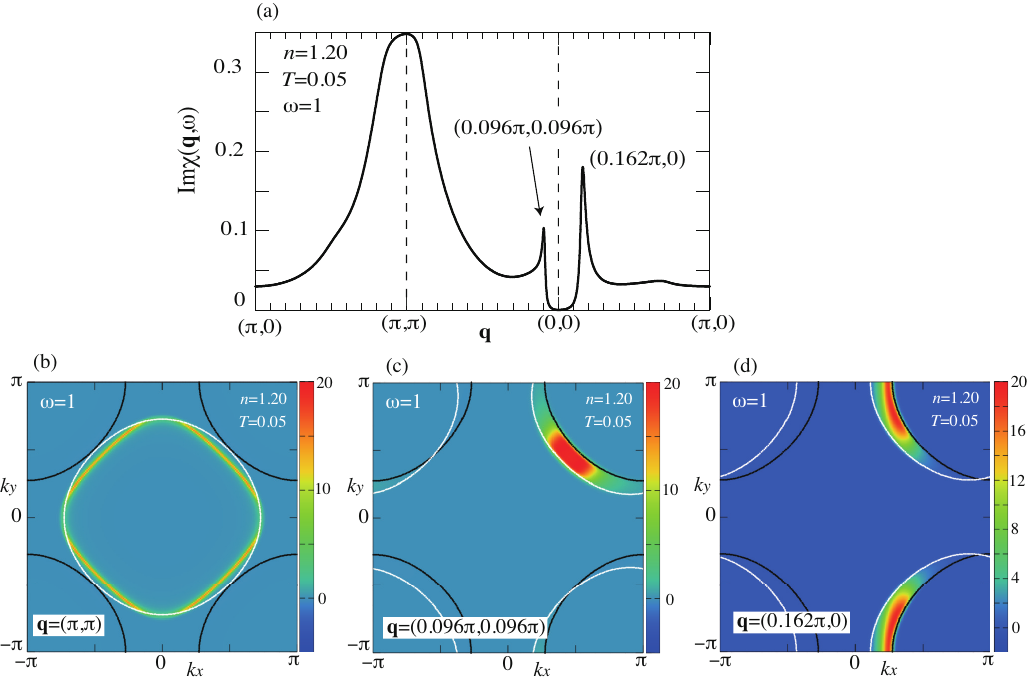}
\caption{Understanding of the structure of Im$\chi(\vq, \omega)$. (a) $\vq$ dependence of Im$\chi(\vq, \omega)$ at $\omega=1$ in the normal state at $T=0.05$ and $n=1.20$. (b)-(d) Maps of \eq{xqw-integrant} in  the Brillouin zone at different $\vq$ for $\omega=1$, $T=0.05$, and $\Gamma=0.1$. The solid curve is the Fermi surface ($\xi_{\vk}=0$) and the white curve is the Fermi surface shifted by $-\vq$ and fulfills $\xi_{\vk+\vq}=0$.}
\label{xqw-process}
\end{figure}

To analyze the magnetic excitation spectrum more closely, we plot the $\vq$ dependence of Im$\chi(\vq, \omega)$ at $\omega=1$ in \fig{xqw-process}(a). There are three characteristic features: a dominant peak around $\vq=(\pi,\pi)$ and sharp peaks at $\vq=(0.096\pi, 0.096\pi)$ and $(0.162\pi, 0)$ close to the upper boundary of the continuum. To understand these features, we present a $\vk$ map of the imaginary part of the integrant in \eq{chi0} after putting $\Delta_{0}=0$ and shifting $\vk$ by $\vq/2$, namely 
\be
 \left( \tanh \frac{\xi_{\vk+\vq}}{2T} - \tanh \frac{\xi_{\vk}}{2T} \right) 
\frac{\Gamma}{ (\xi_{\vk} +\omega - \xi_{\vk+\vq})^{2}  +\Gamma^{2}}  \,,
\label{xqw-integrant}
\ee
in Figs.~\ref{xqw-process}(b)-(d) for different choices of $\vq$ at $\omega=1$. We also superpose the Fermi surface there. For $\vq=(\pi, \pi)$ in \fig{xqw-process}(b), the dominant contributions come from electrons around the nodal regions---$\vk=(\pm\pi/2, \pm \pi/2)$---near the Fermi surface. Given a rather large energy such as $\omega=1$, it is  surprising that states near the Fermi surface are still important at such a high energy. Similarly electrons near the Fermi surface are responsible to form a peak structure at $\vq=(0.096\pi, 0.096\pi)$  and $(0.162\pi, 0)$ as shown in Figs.~\ref{xqw-process}(c) and (d), respectively.

\begin{figure}[b]
\centering 
\includegraphics[width=8cm]{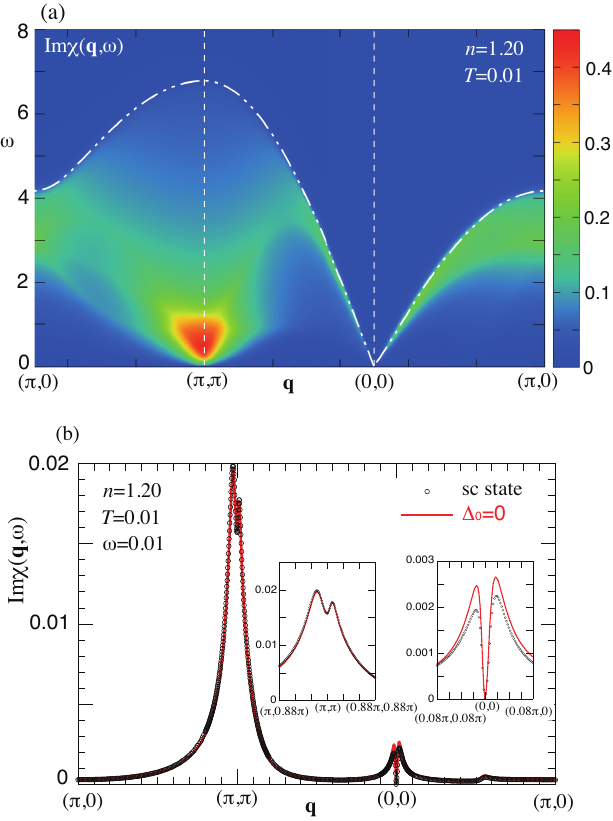}
\caption{(a) $\vq$-$\omega$ map of Im$\chi(\vq, \omega)$ along the symmetry direction at $n=1.20$ in the superconducting state ($T=0.01$). The dashed double-dotted curve is the upper boundary of the continuum spectrum. (b)  Im$\chi(\vq, \omega)$ at $\omega=0.01$ in the superconducting state. The result is compared with that obtained for $\Delta_{0}=0$, but otherwise in the same condition. Regions around $\vq=(\pi, \pi)$ and $(0,0)$ are magnified in the insets.}
\label{xqw-sc}
\end{figure}

To study the effect of superconducting gap, we decrease temperature down to $T=0.01$ and compute Im$\chi(\vq, \omega)$ in the $d$-wave superconducting state in \fig{xqw-sc}(a). No appreciable change is visible and the result is very similar to \fig{xqw-mapT005}. This is mainly due to a small superconducting gap $\Delta_{0} \approx 0.02$, which is much smaller than the energy scale in \fig{xqw-sc}. In the energy range lower than $\Delta_{0}$, however, the effect of the superconducting gap is still tiny as shown in \fig{xqw-sc}(b), where we compare the results with those obtained by putting $\Delta_{0}=0$, but otherwise in the same condition---a visible effect is recognized only around $\vq=(0,0)$. This weak effect of the superconducting gap is traced back to the shape of the Fermi surface shown in \fig{phase}. The low-energy scattering processes with momentum transfer $(\pi, \pi)$ occur around $\vk=(\pm \pi/2,  \pm \pi/2)$, where the $d$-wave superconducting gap has nodes, leading to a weak effect of the superconductivity on the spin excitations. While the strong spectral weight is centered around $\vq=(\pi, \pi)$ in the low-energy region in \fig{xqw-sc}(a), it is interesting that the peak eventually exhibits a small split around $\vq=(\pi,\pi)$ in the low-energy limit [see the inset in \fig{xqw-sc}(b)], although the spectral intensity becomes very small there and finally vanishes in the limit of $\omega=0$.

\begin{figure}[th]
\centering 
\includegraphics[width=8cm]{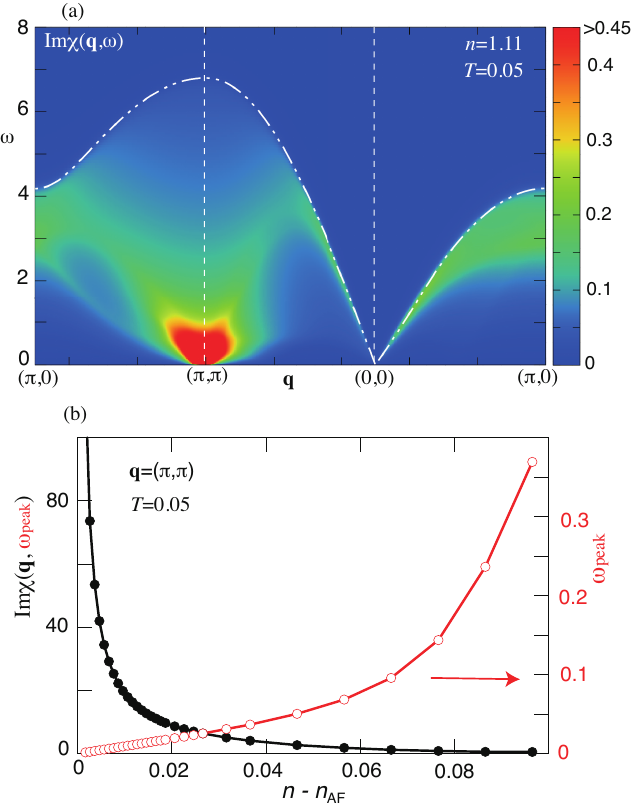}
\caption{(a) $\vq$-$\omega$ map of Im$\chi(\vq,\omega)$ at $n=1.11$ and $T=0.05$ along the symmetry axis near the antiferromagnetic phase. The magnitude of Im$\chi(\vq,\omega)$ more than 0.45 is plotted in the same color, which may make it easy to compare with \fig{xqw-mapT005}. The dashed double-dotted curve is the upper boundary of the continuum spectrum. (b) Peak intensity of Im$\chi(\vq,\omega)$ as a function of the electron density measured from the antiferromagnetic phase ($n_{\rm AF}=1.103$); the peak energy $\omega_{\rm peak}$ is also plotted on the right hand axis. The broadening $\Gamma=0.001$ is taken to capture the criticality correctly. }  
\label{xqw-af}
\end{figure}

How are the magnetic excitation spectra modified upon approaching the magnetic phase? Figure~\ref{xqw-af}(a) demonstrates that the spectral weight increases substantially only around $\vq=(\pi, \pi)$ close to zero energy and the other region does not change much even quantitatively. Here we present the intensity above 0.45 in the same color scale to highlight the effect of the proximity to the magnetic phase through a direct comparison with \fig{xqw-mapT005}. To see the change of the intensity, we plot the peak energy of Im$\chi(\vq, \omega)$ at $\vq=(\pi, \pi)$, which we define as $\omega_{\rm peak}$, and its intensity as a function of the distance from the antiferromagnetic critical point  $n_{\rm AF}$ in \fig{xqw-af}(b). The peak position decreases linearly upon approaching $n_{\rm AF}$ and vanishes at $n=n_{\rm AF}$. At the same time, Im$\chi(\vQ, \omega_{\rm peak})$ diverges as $(n-n_{\rm AF})^{-1}$ in the RPA. 

\begin{figure}[t]
\centering 
\includegraphics[width=8cm]{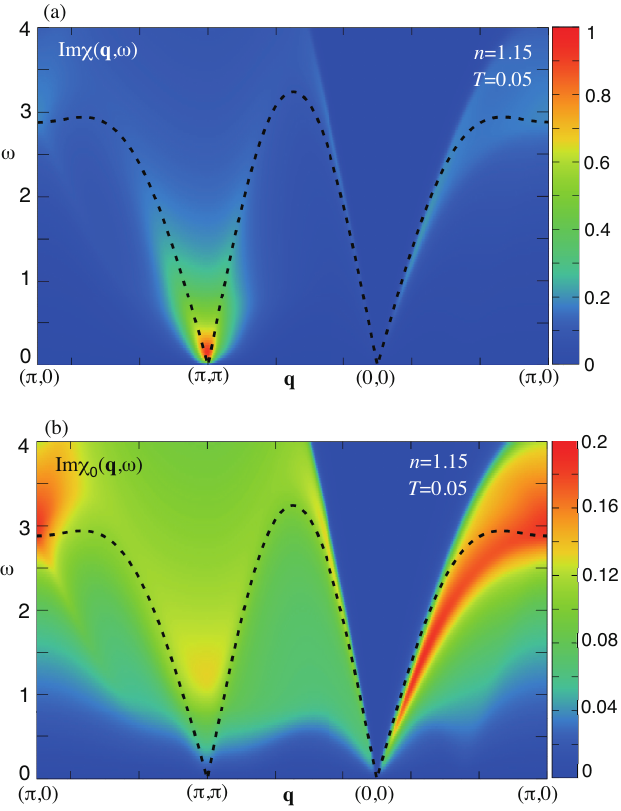}
\caption{Comparison of the spin excitation spectrum at $n=1.15$ and $T=0.05$ with the spin-wave dispersion (dashed curve): (a) Im$\chi(\vq, \omega)$ and (b) Im$\chi_{0}(\vq, \omega)$.  The spin-wave dispersion is described by \eq{spin-wave} with $J=V_{m}$, $J'=0.1J$, $J''=J_{c}=0$, and $Z=1.8$.}
\label{xqw-sw}
\end{figure}

Near the antiferromagnetic phase, one would expect the spin-wave-like dispersion such as
\bea
&&\omega \approx 2Z \sqrt{A(\vq)^{2} - B(\vq)^{2}} 
\label{spin-wave}  \,, \\
&&A(\vq)=J- J_{c}/2 - (J'- J_{c}/4)(1- \cos q_{x} \cos q_{y}) - J''[1 - (\cos 2 q_{x} + \cos 2 q_{y})/2] \,, \\
&&B(\vq)=(J-J_{c}/2)(\cos q_{x} + \cos q_{y})/2 \,,
\eea
where $J$, $J'$, and $J''$ are the first, second, and third nearest-neighbor spin exchange interactions on a square lattice and $J_{c}$ is the ring exchange interaction among four spins on a square plaquette; $Z$ is a renormalization factor \cite{coldea01}.  We superpose the dispersion \eq{spin-wave} by taking $J=V_{m}=1$, $J'=0.1J$, $J''=J_{c}=0$, and $Z=1.8$ in \fig{xqw-sw}. A surprise here is that the dispersive feature of Im$\chi(\vq, \omega)$ is well characterized by the spin-wave dispersion even far away from $\vq=(\pi, \pi)$, actually in the entire momentum region, although calculations are performed in the paramagnetic phase. To check the effect of collective fluctuations, we present a $\vq$-$\omega$ map of Im$\chi_{0}(\vq, \omega)$ for the same parameters as those in \fig{xqw-sw}(a) and superpose the same spin-wave dispersion. A big discrepancy is recognized around $\vq=(\pi, \pi)$ in $\omega \lesssim 2$, indicating the collective magnetic fluctuations there.  Except for such a region, \fig{xqw-sw}(b) shows that the spectrum of Im$\chi_{0}(\vq, \omega)$ is also captured by the spin-wave dispersion. This means that a high-energy region and a low-energy regions around $\vq=(0, 0)$ in \fig{xqw-sw}(a) is characterized mainly by individual particle-hole excitations, not collective excitations such as paramagnons.

\begin{figure}[hb]
\centering 
\includegraphics[width=6cm]{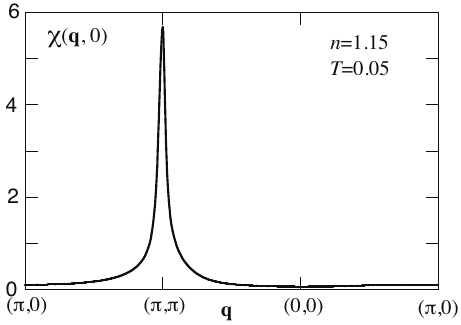}
\caption{Static part of the magnetic susceptibility $\chi(\vq,0)$ at $n=1.15$ and $T=0.05$.} 
\label{xq}
\end{figure}

While it is interesting to recognize that the spectrum of Im$\chi(\vq, \omega)$ in \fig{xqw-sw}(a) can be fitted in terms of the spin-wave dispersion, it does not indicate collective features of the magnetic excitations such as paramagnons at least around $\vq=(0,0)$. Rather the dispersive feature of Im$\chi(\vq, \omega)$ around $\vq=(0,0)$ is characterized by the boundary of the individual particle-hole excitations. We also point out that it is a weak feature in the sense that the static part $\chi(\vq, 0)$ exhibits no pronounced structure around $\vq=(0,0)$, but only a dominant peak around $\vq=(\pi, \pi)$ as shown in \fig{xq}.

\subsection{Bond-charge excitation spectrum}
Next we study bond-charge excitations. Since the dynamical bond-charge susceptibility $\hat{\kappa}(\vq, \omega)$ in \eq{kappaRPA} is described by the $4 \times 4$ symmetric matrix, we obtain 10 different bond-charge susceptibilities. The physical ones are given by the diagonal components and thus we focus on them---see Appendix~A for spectra of the off-diagonal components.

\begin{figure}[b]
\centering 
\includegraphics[width=16cm]{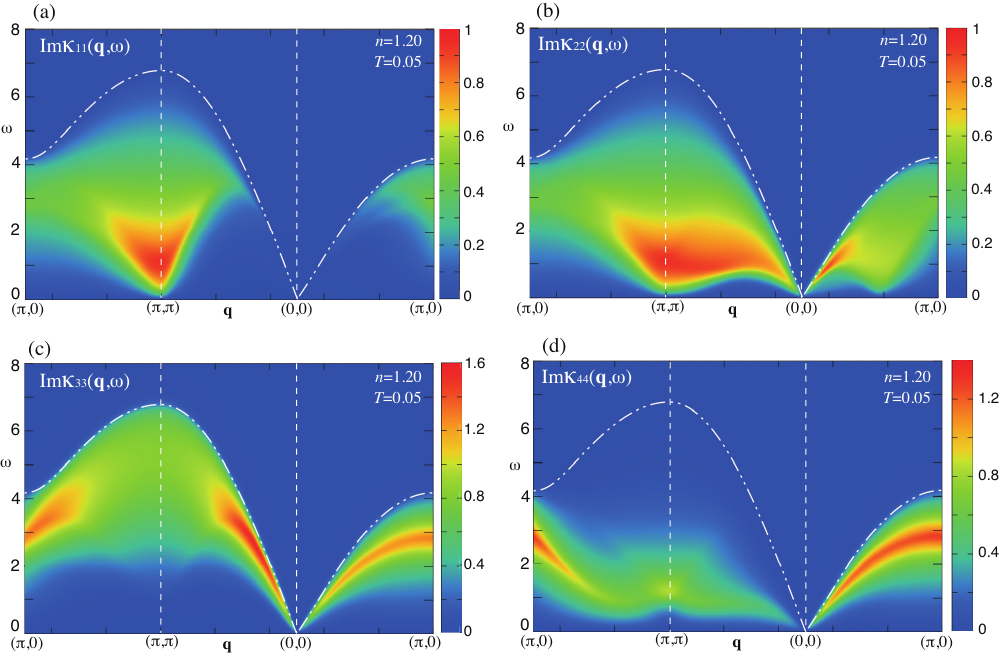}
\caption{$\vq$-$\omega$ maps of the diagonal components of the bond-charge excitation spectrum along the $(\pi,0)$-$(\pi,\pi)$-$(0,0)$-$(\pi,0)$ direction at $n=1.20$ in the normal state ($T=0.05$): (a) Im$\kappa_{11}(\vq, \omega)$, (b) Im$\kappa_{22}(\vq, \omega)$, (c) Im$\kappa_{33}(\vq, \omega)$, and (d) Im$\kappa_{44}(\vq, \omega)$. The dashed double-dotted curve is the upper boundary of the continuum spectrum.}
\label{kqw-mapT005}
\end{figure}

\begin{figure}[tb]
\centering 
\includegraphics[width=16cm]{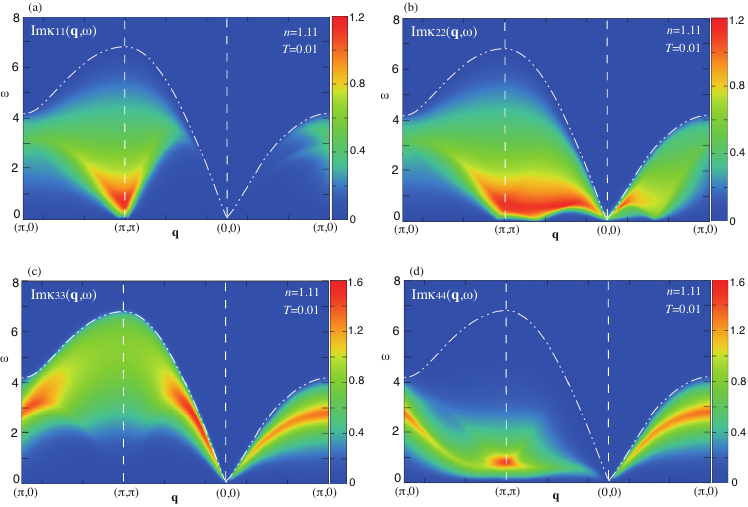}
\caption{$\vq$-$\omega$ maps of the diagonal components of the bond-charge excitation spectrum along the $(\pi,0)$-$(\pi,\pi)$-$(0,0)$-$(\pi,0)$ direction at $n=1.11$ in the superconducting state ($T=0.01$): (a) Im$\kappa_{11}(\vq, \omega)$, (b) Im$\kappa_{22}(\vq, \omega)$, (c) Im$\kappa_{33}(\vq, \omega)$, and (d) Im$\kappa_{44}(\vq, \omega)$. The dashed double-dotted curve is the upper boundary of the continuum spectrum.} 
\label{kqw-sc}
\end{figure}

\begin{figure}[bht]
\centering 
\includegraphics[width=10cm]{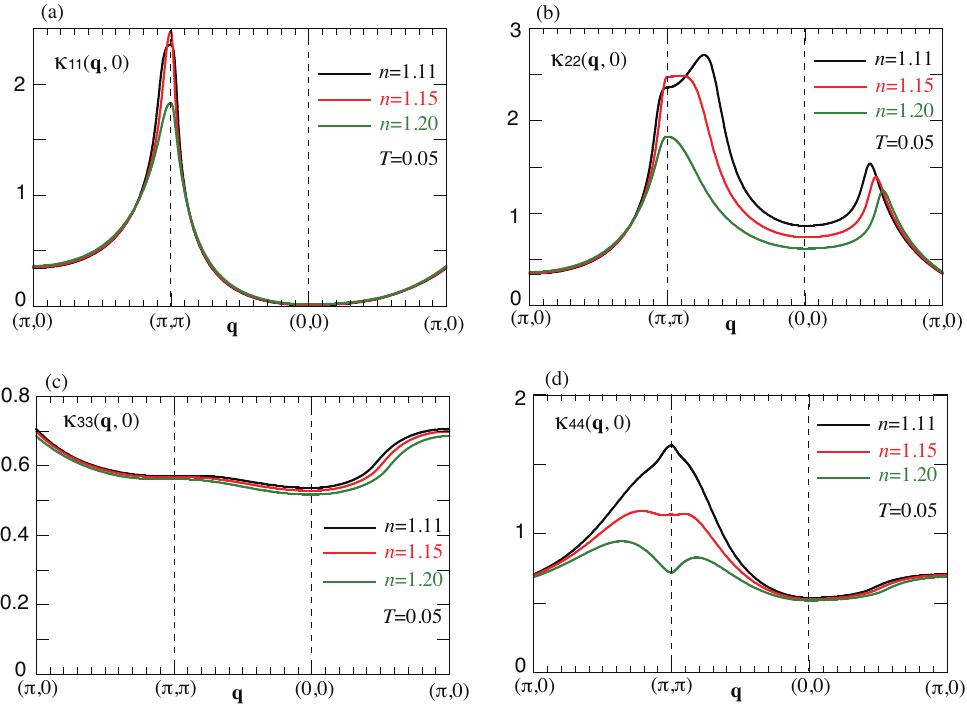}
\caption{$\vq$ dependence of the static bond-charge susceptibilities, (a) $\kappa_{11}(\vq,0)$, (b) $\kappa_{22}(\vq,0)$, (c) $\kappa_{33}(\vq,0)$, and (d) $\kappa_{44}(\vq,0)$, along the $(\pi,0)$-$(\pi,\pi)$-$(0,0)$-$(\pi,0)$ direction in the normal state ($T=0.05$) for three choices of the electron density $n=1.11$, $1.15$, and $1.20$. }
\label{kq-n}
\end{figure}

Figure~\ref{kqw-mapT005}(a) shows a $\vq$-$\omega$ map of Im$\kappa_{11}(\vq, \omega)$, namely $s$-wave bond-charge excitations. The overall feature is similar to the magnetic excitation spectrum shown in \fig{xqw-mapT005} and the spectral weight is centered around $\vq=(\pi, \pi)$. The $d$-wave bond-charge excitations [\fig{kqw-mapT005}(b)] also have large spectral weight around $\vq=(\pi, \pi)$, but in contrast to the $s$-wave ones, extends largely toward $\vq=(0,0)$. A characteristic feature of the $d$-wave bond-charge excitations is softening near $\vq=(0.5 \pi, 0)$, where the spectral weight extends to zero energy, reminiscent of a kind of the Kohn anomaly \cite{gruener88}. Figure~\ref{kqw-mapT005}(c) is the spectrum of the bond charge with the form factor $p^{+}_{\vk}$  [\eq{formfactor}] and is distinct from Figs.~\ref{kqw-mapT005}(a) and (b)---there is no strong spectral weight around $\vq=(\pi, \pi)$ in low energy. Instead Im$\kappa_{33}(\vq\, \omega)$ exhibits a dispersive feature close to the boundary of the particle-hole excitations especially along the directions $\vq=(0,0)$-$(\pi,\pi)$ and $(0,0)$-$(\pi,0)$. These are, however, not collective modes, but a peak structure of the particle-hole continuum. The bond-charge excitations with the form factor $p^{-}_{\vk}$  [\eq{formfactor}] is shown in \fig{kqw-mapT005}(d). There is a dispersive feature in the direction of $\vq=(0,0)$-$(\pi,0)$, which is also a peak structure of the particle-hole excitations and should not be associated with a collective mode. Rather, collective excitations are realized around $\vq=(\pi,\pi)$ and $\omega\sim 1$, which becomes sharper at lower energy upon approaching the antiferromagnetic phase as we shall see later [see Figs.~\ref{kqw-sc}(d) and \ref{kq-n}(d)]. 

\begin{figure}[bht]
\centering 
\includegraphics[width=10cm]{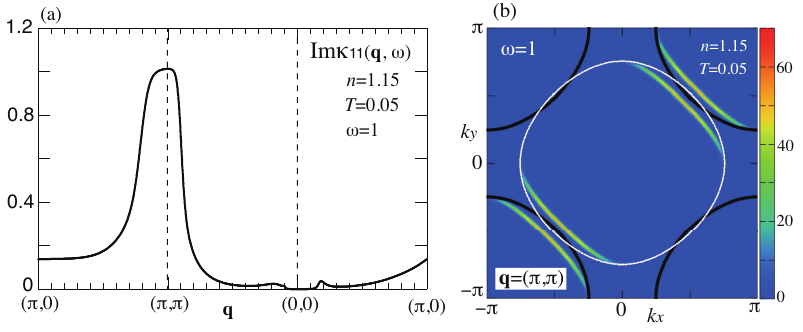}
\caption{Understanding of the major structure of Im$\kappa_{11}(\vq, \omega)$. (a) $\vq$ dependence of Im$\kappa_{11}(\vq, \omega)$ at $\omega=1$ in the normal state at $T=0.05$ and $n=1.15$. (b) Map of \eq{integrant-kqw} in  the Brillouin zone for $\vq=(\pi, \pi)$, $\omega=1$, $T=0.05$, and $\Gamma=0.1$ with $\gamma_{\vk_{+}}=s_{\vk_{+}}$. The solid curve is the Fermi surface ($\xi_{\vk}=0$) and the white curve is the Fermi surface shifted by $-\vq$ and fulfills $\xi_{\vk+\vq}=0$. 
}
\label{kqw11-process}
\end{figure}

To understand the characteristic feature of each bond-charge excitation spectrum shown in \fig{kqw-mapT005}, we present the $\vq$ dependence of Im$\kappa_{11}(\vq, \omega)$ at $\omega=1$ in \fig{kqw11-process}(a), showing a peak around $\vq=(\pi, \pi)$  as in the case of the magnetic excitations [\fig{xqw-process}(a)]. In \fig{kqw11-process}(b), we present the integrant of the Im$\kappa_{0\, 11}(\vq, \omega)$ [see \eq{kqw1122}] in the normal phase, which is given by 
\be
\gamma_{\vk_{+}}^{2} \left(\tanh \frac{\xi_{\vk}}{2T} - \tanh \frac{\xi_{\vk+\vq}}{2T}\right) 
\left[ 
\frac{-\Gamma}{ (\xi_{\vk}+ \omega - \xi_{\vk+\vq})^{2}  + \Gamma^{2} } +
\frac{\Gamma}{ (\xi_{\vk} - \omega - \xi_{\vk+\vq})^{2}  + \Gamma^{2} } \right]
\label{integrant-kqw}
\ee
after shifting $\vk$ by $\vq/2$; $\gamma_{\vk_{+}}=s_{\vk_{+}}$ [see \eq{formfactor-s}]. For convenience we also plot the Fermi  surfaces given by $\xi_{\vk}=0$ (solid curve) and $\xi_{\vk+\vq}=0$ (white curve). Although we take energy as high as $\omega=1$, the major scattering still occurs in regions near the Fermi surface. These regions are different from those in the magnetic excitations [\fig{xqw-process}(b)] because of the presence of the form factor $s_{\vk_{+}}$. Nonetheless, it is  interesting that a similar peak at $\vq=(\pi, \pi)$ is realized in both cases. 

\begin{figure}[b]
\centering 
\includegraphics[width=10cm]{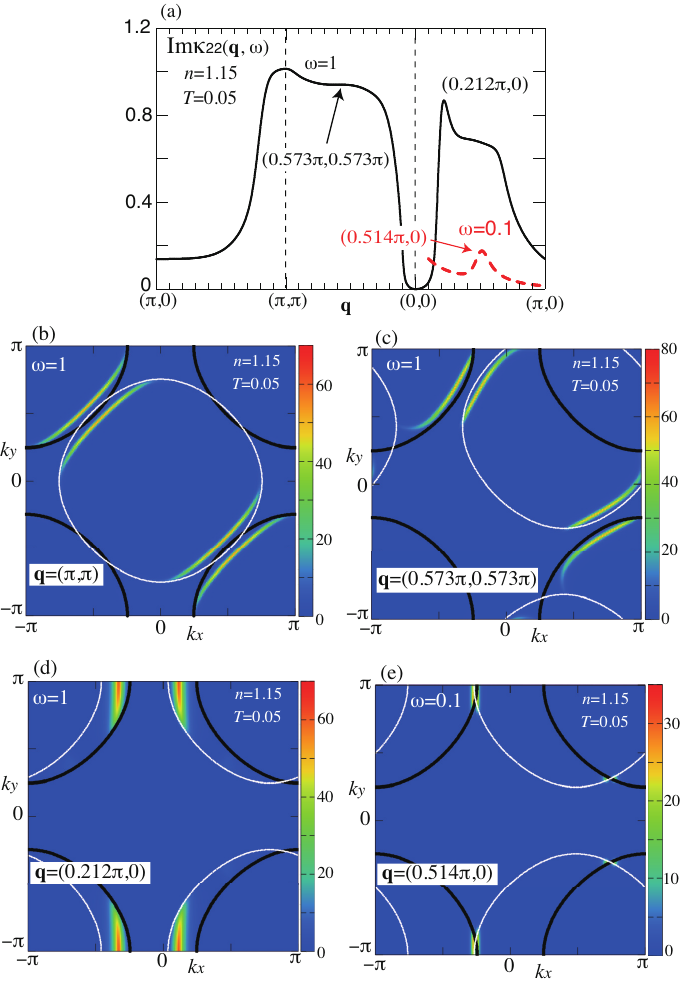}
\caption{Understanding of the structure of Im$\kappa_{22}(\vq, \omega)$. (a) $\vq$ dependence of Im$\kappa_{22}(\vq, \omega)$ at $\omega=1$ in the normal state at $T=0.05$ and $n=1.15$. Results at $\omega=0.1$ are also presented along the direction of $(0,0)$-$(\pi,0)$ with a dashed red curve. (b)-(d) Maps  of \eq{integrant-kqw} in  the Brillouin zone for different $\vq$ at $\omega=1$, $T=0.05$, and $\Gamma=0.1$ with $\gamma_{\vk_{+}}=d_{\vk_{+}}$. The solid curve is the Fermi surface ($\xi_{\vk}=0$) and the white curve is the Fermi surface shifted by $-\vq$ and fulfills $\xi_{\vk+\vq}=0$. (e) The same as (d) except that $\omega=0.1$ and $\vq=(0.514\pi,0)$ are taken. 
}
\label{kqw22-process}
\end{figure}

We next analyze the $d$-wave bond-charge excitations in a similar way. There is a very broad spectrum in the direction $(\pi, \pi)$-$(0, 0)$ in \fig{kqw22-process}(a) and the characteristic momenta may be given by $\vq=(\pi, \pi)$ and $(0.573\pi, 0.573\pi)$. For each momentum we plot in \fig{kqw22-process}(b) and (c) a map of the integrant of Im$\kappa_{0\, 22}(\vq, \omega)$ in the normal state, i.e., \eq{integrant-kqw} with $\gamma_{\vk_{+}}=d_{\vk_{+}}$ together with the Fermi surface. At $\vq=(\pi, \pi)$, the momentum region where the integrant becomes large is obtained by reflecting the intensity map of the $s$-wave case [\fig{kqw11-process}(b)] with respect to $k_{x}$ axis. At $\vq=(0.573\pi, 0.573\pi)$, on the other hand, the integrant is enhanced in regions around $\vk=(\pi, 0)$ and $(0,\pi)$, extending toward nodal regions. While the energy $\omega=1$ is rather large, electrons near the Fermi surface are responsible to create the structure seen in \fig{kqw22-process}(a). A sharp feature of Im$\kappa_{22}(\vq, \omega)$ is realized at $\vq=(0.212\pi, 0)$ in \fig{kqw22-process}(a), which is driven by electrons near the Fermi surface close to $(0, \pm\pi)$ as seen in \fig{kqw22-process}(d).

The $d$-wave bond-charge excitations have a special feature to exhibit softening near $\vq=(0.5\pi, 0)$ at low $\omega$ as shown in \fig{kqw-mapT005}(b). To understand such a feature, we also present the $\vq$ dependence of Im$\kappa_{22}(\vq, \omega)$ at $\omega=0.1$ in the direction of $\vq=(0,0)$-$(\pi,0)$ in \fig{kqw22-process}(a) and a  map of the integrant \eq{integrant-kqw}  in \fig{kqw22-process}(e). This clearly shows that electrons in the vicinity of the Fermi surface near $\vk=(0,\pm\pi)$ are responsible for the softening of the $d$-wave bond-charge excitations near $\vq=(0.5\pi, 0)$ in \fig{kqw-mapT005}(b).

\begin{figure}[thb]
\centering 
\includegraphics[width=10cm]{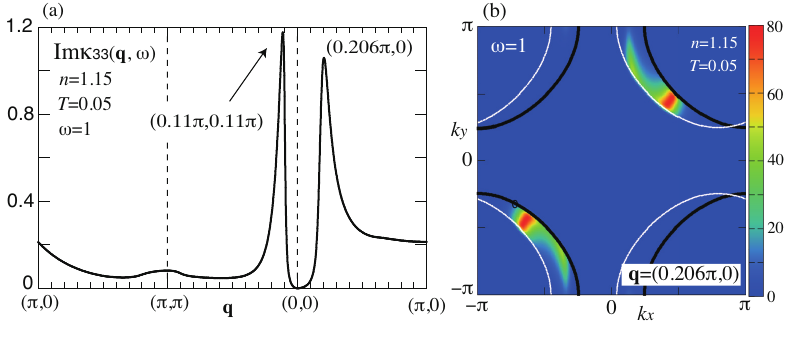}
\caption{Understanding of the structure of Im$\kappa_{33}(\vq, \omega)$. (a) $\vq$ dependence of Im$\kappa_{33}(\vq, \omega)$ at $\omega=1$ in the normal state at $T=0.05$ and $n=1.15$.  (b) Map of \eq{integrant-kqw} in  the Brillouin zone for $\vq=(0.206\pi, 0)$, $\omega=1$, $T=0.05$, and $\Gamma=0.1$ with $\gamma_{\vk_{+}}=p_{\vk_{+}}^{+}$. The solid curve is the Fermi surface ($\xi_{\vk}=0$) and the white curve is the Fermi surface shifted by $-\vq$ and fulfills $\xi_{\vk+\vq}=0$. 
}
\label{kqw33-process}
\end{figure}

Im$\kappa_{33}(\vq, \omega)$ exhibits sharp peaks  around $\vq=(0,0)$ as shown in \fig{kqw33-process}(a). The integrant of Im$\kappa_{0\, 33}(\vq, \omega)$ in the normal state is given by the same expression as \eq{integrant-kqw} by taking $\gamma_{\vk_{+}}=p_{\vk_{+}}^{+}$ [see \eq{formfactor}], although it becomes different in the superconducting state; compare \eq{kqw3344} with \eq{kqw1122}. For $\vq=(0.206\pi, 0)$ the integrant is enhanced around $\vk=(\pm\pi/2, \pm\pi/2)$ near the Fermi surface. A result corresponding to a map for $\vq=(0.11\pi, 0.11\pi)$ is similar to \fig{kqw33-process}(b), which may be easily expected by recognizing the momentum $\vq$ is close to each other.

\begin{figure}[bt]
\centering 
\includegraphics[width=10cm]{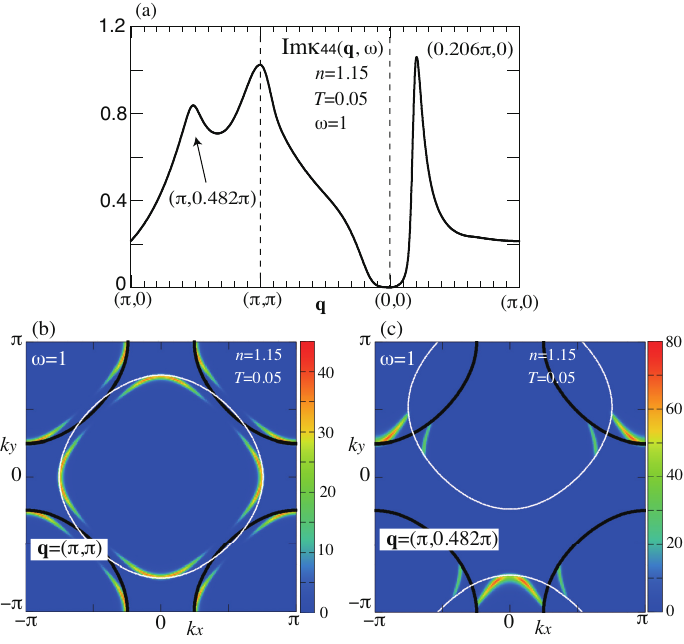}
\caption{Understanding of the structure of Im$\kappa_{44}(\vq, \omega)$. (a) $\vq$ dependence of Im$\kappa_{44}(\vq, \omega)$ at $\omega=1$ in the normal state at $T=0.05$ and $n=1.15$.  (b) and (c) Maps of \eq{integrant-kqw} in  the Brillouin zone for $\vq=(\pi, \pi)$ and $(\pi, 0.482\pi)$, respectively, at $\omega=1$, $T=0.05$, and $\Gamma=0.1$ with $\gamma_{\vk_{+}}=p_{\vk_{+}}^{-}$. The solid curve is the Fermi surface ($\xi_{\vk}=0$) and the white curve is the Fermi surface shifted by $-\vq$ and fulfills $\xi_{\vk+\vq}=0$. 
}
\label{kqw44-process}
\end{figure}

The $\vq$ dependence of Im$\kappa_{44}(\vq, \omega)$ at $\omega=1$ is shown in \fig{kqw44-process}(a). There are three characteristic peaks at $\vq=(\pi, 0.482\pi)$, $(\pi, \pi)$, and $(0.206\pi, 0)$. The last momentum is the same as that shown in \fig{kqw33-process}. This is because $\kappa_{0 \, 33}((q_{x}, 0), iq_{m})=\kappa_{0\, 44}((q_{x}, 0), iq_{m})$ as is easily seen by replacing $k_{y} \rightarrow - k_{y}$ on the right hand side in the $\vk$ summation in \eq{kqw3344}. Hence the integrant has a large value around $\vk=(\mp \pi/2, \pm \pi /2)$ by replacing $k_{y} \rightarrow - k_{y}$ in \fig{kqw33-process}(b). For $\vq=(\pi, \pi)$, \fig{kqw44-process}(b) shows that the integrant [\eq{integrant-kqw} with $\gamma_{\vk_{+}}=p_{\vk_{+}}^{-}$] has a large value near $\vk=(\pm \pi, 0)$ and $(0,\pm \pi)$, which is in sharp contrast to the case of $s$- and $d$-wave bond-charge excitations at $\vq=(\pi, \pi)$ shown in Figs.~\ref{kqw11-process}(b) and \ref{kqw22-process}(b). The importance of electrons near the Fermi surface, however, is shared with other bond-charge excitations even for $\omega=1$. For $\vq=(\pi, 0.482\pi)$, the integrant is enhanced around $\vk=(\pm \pi, 0)$ and $(0, -\pi)$  near the Fermi surface, with keeping a symmetry with respect to $k_{x} \rightleftarrows -k_{x}$.

Next we study $\vq$-$\omega$ maps of the bond-charge excitations by decreasing the doping and temperature in the superconducting state. Figure~\ref{kqw-sc}  summarizes results at $n=1.11$ and $T=0.01$. A comparison between \fig{kqw-sc} and \ref{kqw-mapT005} reveals that the doping and temperature dependences are very mild and qualitative features do not change. Quantitatively the spectral weight of the $d$-wave bond-charge excitations is substantially enhanced at low energy in a wide region from $\vq=(\pi, \pi)$ to $(\pi/2,\pi/2)$ [compare \fig{kqw-sc}(b) with \fig{kqw-mapT005}(b)] and Im$\kappa_{44}(\vq, \omega)$ has stronger intensity around $\vq=(\pi, \pi)$ at slightly lower energy [compare \fig{kqw-sc}(d) with \fig{kqw-mapT005}(d)]. The obtained results of Figs.~\ref{kqw-sc}(b) and (d) might be counterintuitive because one would expect a sizable effect of the superconducting gap if one recalls that electrons near the Fermi surface around $(\pi, 0)$ and $(0, \pi)$ in \fig{kqw22-process}(c) and  $(\pm \pi, 0)$ and $(0, \pm \pi)$ in \fig{kqw44-process}(b) are responsible to form a structure at $\vq=(0.573\pi, 0.573\pi)$ and $(\pi, \pi)$, respectively, and the superconducting gap is characterized by $d$-wave symmetry that yields the largest gap around  $\vk=(\pm \pi, 0)$ and $(0, \pm \pi)$. However, this is simply due to a quantitative aspect that the largest gap of the superconductivity is around $\Delta_{0}=0.055$, much smaller energy than that in \fig{kqw-sc}.

\begin{figure}[th]
\centering 
\includegraphics[width=10cm]{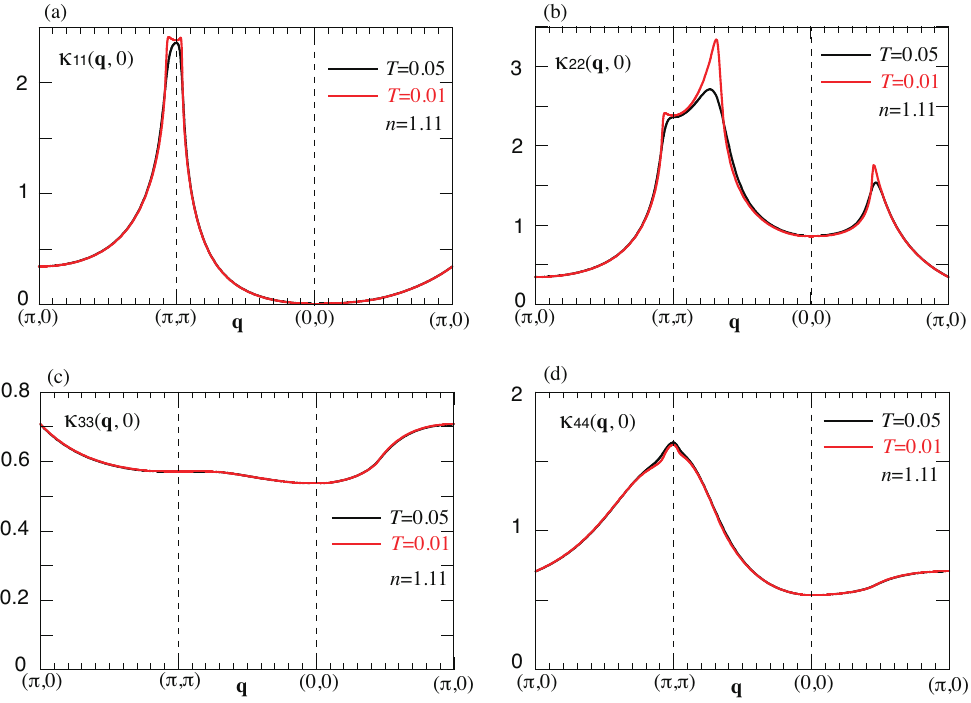}
\caption{$\vq$ dependence of the static bond-charge susceptibilities, (a) $\kappa_{11}(\vq,0)$, (b) $\kappa_{22}(\vq,0)$, (c) $\kappa_{33}(\vq,0)$, and (d) $\kappa_{44}(\vq,0)$, along the $(\pi,0)$-$(\pi,\pi)$-$(0,0)$-$(\pi,0)$ direction at $n=1.11$ in the normal state at two choices of temperatures $T=0.05$ and $0.01$; $\Delta_{0}=0$ is assumed at $T=0.01$ for a comparison.} 
\label{kq-T}
\end{figure}

The temperature and doping dependences of the bond-charge excitations can be studied more clearly by considering the static part of the bond-charge susceptibility. Figure~\ref{kq-T} shows the $\vq$ dependence of $\kappa_{ii}(\vq, 0)$ at $T=0.01$ and $0.05$ for $n=1.11$. A clear temperature dependence is recognized only for the $d$-wave bond-charge susceptibility, indicating the development of a sharp peak at $\vq=(0.688\pi, 0.688 \pi)$. Except for this, the temperature dependence is weak. 

The static bond-charge susceptibilities shown in \fig{kq-T} also indicate clearly the characteristic momenta of each susceptibility. The $s$-wave bond-charge susceptibility is characterized by $\vq=(\pi, \pi)$, the $d$-wave one by $\vq=(\pi, \pi)$, $(0.688\pi, 0.688\pi)$, and $(0.456\pi, 0)$, $\kappa_{33}(\vq, 0)$ is structureless, and $\kappa_{44}(\vq, 0)$ has a peaks at $\vq=(\pi, \pi)$. 

The doping dependence of the static susceptibilities is shown in \fig{kq-n} by plotting their $\vq$ dependence for three choices of the electron density.  The $s$-wave bond-charge susceptibility has a sharper peak at $\vq=(\pi, \pi)$ for  lower electron density, but does not exhibit the tendency of the divergence in \fig{kq-n}(a). The $d$-wave part shown in \fig{kq-n}(b) tends to have an incommensurate peak away from $(\pi, \pi)$ with decreasing the electron density. At the same time,  the peak along the direction $(0,0)$-$(\pi, 0)$ is enhanced gradually with decreasing the electron density. While $\kappa_{33}(\vq, \omega)$ is structureless independent of the electron density [\fig{kq-n}(c)], $\kappa_{44}(\vq, 0)$ develops a commensurate peak at $\vq=(\pi, \pi)$ with decreasing the electron density as shown in \fig{kq-n}(d).

\subsection{Extension to hole-doped side}
So far we have focused on the electron-doped side and we shall discuss later that the obtained results capture essential features observed in electron-doped cuprates. It is a natural question how results change if we consider the hole-doped side. We therefore extend the present study by taking the electron density less than $n=1$, otherwise, keeping the same parameters, aiming to clarify similarities and differences between the electron-doped and hole-doped sides.

\subsubsection{Phase diagram}

\begin{figure}[thb]
\centering 
\includegraphics[width=8cm]{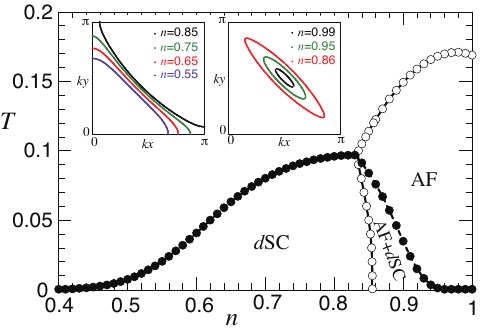}
\caption{Phase diagram on the hole-doped side. It is obtained by extending the electron density to the region $n \leq 1$ in \fig{phase}---otherwise the same parameters are used. Two insets are Fermi surfaces obtained by assuming $\Delta_{0}=0$ at $T=0.01$ for several choices of the electron density. The large Fermi surface is reconstructed in the AF phase to form hole pockets around $\vk=(\pi/2, \pi/2)$ in $n \geq 0.86$. 
}
\label{phase-hole}
\end{figure}

Figure~\ref{phase-hole} shows the phase diagram in the plane of the electron density $(n \leq 1)$ and temperature obtained  in the present model. Similar to \fig{phase}, there are three phases: antiferromagnetic phase, $d$-wave superconducting phase, and their coexistence---no bond-order instability is found. The antiferromagnetic phase 
extends up to $\sim 15 \%$ hole doping and tends to survive against carrier doping more than the electron-doped side  (\fig{phase}). In sharp contrast to the electron-doped side, the coexistence with superconductivity is less favorable and the superconductivity is strongly suppressed down to zero temperature especially close to half-filling ($n=1)$. In the (pure) superconducting phase, the onset temperature is enhanced by a factor of 2 more than the electron-doped side  and extends up to much higher doping rate. 

The shape of the Fermi surface is also different from the electron-doped side. In the antiferromagnetic phase, hole pockets are realized around $\vk=(\pi/2, \pi/2)$ and elongate along the magnetic Brillouin zone boundary given by $| k_{x} | + | k_{y} | = \pi$ with hole doping. The large hole-like Fermi surface is realized when entering the paramagnetic phase and changes to the electron-like Fermi surface upon further doping. 

\subsubsection{Magnetic excitation spectrum} 

\begin{figure}[b]
\centering 
\includegraphics[width=15cm]{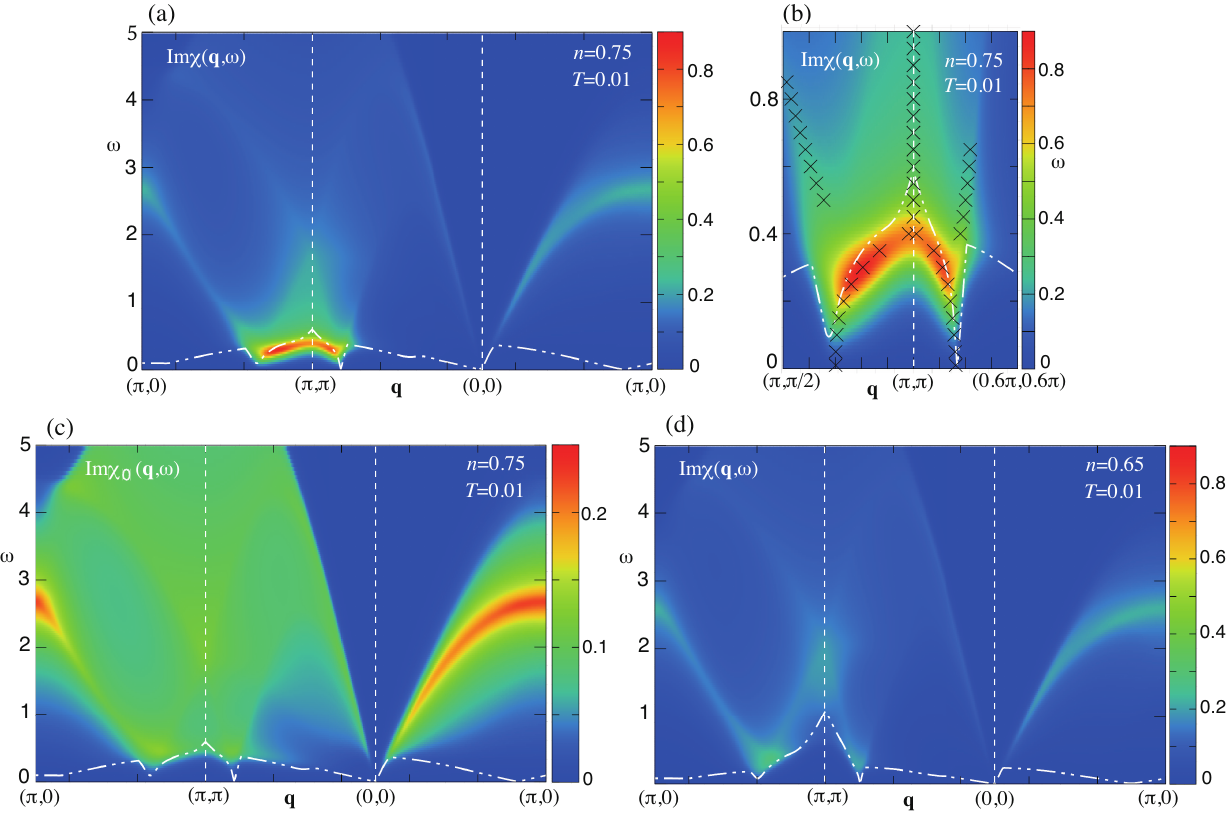}
\caption{$\vq$-$\omega$ maps of Im$\chi(\vq, \omega)$ on the hole-doped side along the direction $(\pi,0)$-$(\pi, \pi)$-$(0,0)$-$(\pi, 0)$ in the superconducting phase ($T=0.01$) at $n=0.75$. The lower edge of the continuum is depicted by the dashed double-dotted curve. (b) Zoom around $\vq=(\pi, \pi)$ in the low-energy region in (a). Crosses specify peak positions at a given $\omega$.  (c) $\vq$-$\omega$ map of Im$\chi_{0}(\vq, \omega)$ for the same parameters as those in (a). (d) $\vq$-$\omega$ map of Im$\chi(\vq, \omega)$ upon further doping ($n=0.65$). The color scale is the same as that in (a) to make a direct comparison with (a) possible. 
}
\label{xqw-map-hole}
\end{figure}

The magnetic excitation spectrum is shown in \fig{xqw-map-hole}(a) along the direction $(\pi,0)$-$(\pi, \pi)$-$(0,0)$-$(\pi, 0)$  in the superconducting phase ($T=0.01$) at $n=0.75$. The gap of the particle-hole excitations is also computed by evaluating the minimal energy of $E_{\vk}+ E_{\vk+\vq}$ for a given $\vq$ when $\vk$ runs in the entire Brillouin zone. A dominant intensity is realized around $\vq=(\pi, \pi)$, similar to the electron-doped side (\fig{xqw-mapT005}), but it appears inside the gap of the particle-hole excitations. This is the spin resonance mode. Its dispersion is highlighted in \fig{xqw-map-hole}(b). It forms the so-called hourglass dispersion. Below the resonance energy $(\omega \leq 0.4)$ at $\vq=(\pi, \pi)$, it forms a downward dispersion and its peak position goes away from $\vq=(\pi, \pi)$, yielding an incommensurate structure. It is interesting to recognize that the strongest intensity is realized at an incommensurate wavevector along the downward dispersion. The resonance mode rapidly damps upon entering the particle-hole continuum, but an upward-like dispersion is still resolved, with retaining large spectral weight at $\vq=(\pi, \pi)$. The presence  of the spin-resonance mode is a consequence of a large superconducting gap and the proximity to the antiferromagnetic phase, which is clearly recognized by comparing \fig{xqw-map-hole}(a) with (c), where Im$\chi_{0}(\vq, \omega)$ is presented. A close comparison with Fig.~\ref{xqw-map-hole}(a) in the higher energy reveals that a remnant of the upward dispersion as well as relatively large weight at $(\pi, \pi)$ is already realized in \fig{xqw-map-hole}(c), indicating that individual particle-hole excitations are largely involved in the upward dispersion. 

In \fig{xqw-map-hole}(a), a steep dispersion is realized around $\vq=(0,0)$ near the boundary of the particle-hole continuum spectrum. As seen in \fig{xqw-map-hole}(c), this dispersion is already realized in Im $\chi_{0}(\vq, \omega)$ and thus should not be associated with any collective mode such as paramagnons. This feature is shared with the electron-doped side (see \fig{xqw-sw}).  

In \fig{xqw-map-hole}(d), we show how the magnetic excitations change with further carrier doping, namely further away from the antiferromagnetic phase. Magnetic correlations become weaker, as expected, and can no longer host the spin resonance mode---a peak structure is seen only inside the continuum spectrum. In particular, the incommensurate peak becomes prominent at $\vq \approx (\pi, 0.6\pi)$, from which two dispersive features develop in high energy toward both $\vq=(\pi, 0)$ and $(\pi, \pi)$. The latter dispersion approaches $\vq=(\pi, \pi)$ with increasing energy  and eventually changes to a board peak at $\vq=(\pi, \pi)$ and $\omega \approx 1.5$. Interestingly, upon further increasing energy ($\omega \gtrsim 2$), we can barley see two dispersive features emerging from $\vq=(\pi, \pi)$ almost in a linear form in momentum. This is a peak of the continuum and does not imply some ``hidden'' mode. In fact, this feature can also be recognized as a faint structure in Im$\chi_{0}(\vq, \omega)$ [\fig{xqw-map-hole}(c)] in $\omega \gtrsim 2$.

In spite of the incommensurate magnetic correlation the antiferromagnetic phase is commensurate, namely characterized by the momentum $\vq=(\pi, \pi)$ in \fig{phase-hole}. This could be easily understood by noting that the static part of the susceptibility is given by the energy integration of the spectrum via Kramers-Kronig relations and thus the static susceptibility can still exhibit the commensurate peak.

The superconducting gap is decreased upon carrier doing as expected from the doping dependence of the onset temperature of superconducting phase  in \fig{phase-hole}. Nonetheless, the gap of the continuum around $\vq=(\pi, \pi)$ at $n=0.65$ is larger than that at $n=0.75$; compare Figs.~\ref{xqw-map-hole}(d) and (c). This larger gap comes from the Fermi surface shape (see the inset of \fig{phase-hole}), which does not allow low-energy particle-hole excitations with a momentum transfer $\vq \approx (\pi, \pi)$ even in the normal state.

\begin{figure}[tb]
\centering 
\includegraphics[width=7cm]{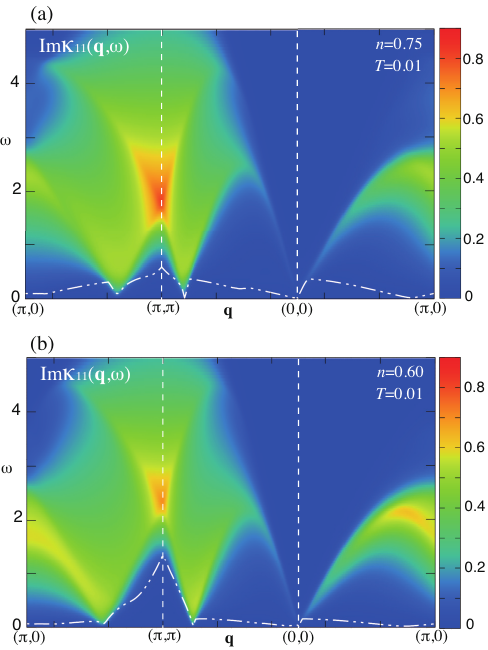}
\caption{$\vq$-$\omega$ maps of Im$\kappa_{11}(\vq, \omega)$ on the hole-doped side along the $(\pi,0)$-$(\pi,\pi)$-$(0,0)$-$(\pi,0)$ direction in the superconducting state ($T=0.01$): (a) $n=0.75$ and (b) $n=0.60$. The lower edge of the continuum is depicted by the dashed double-dotted curve.}
\label{kqw11-map-hole}
\end{figure}

\subsubsection{Bond-charge excitation spectrum} 
Next we clarify bond-charge excitation spectra, which show many features different from those observed in the electron-doped side given in Sec. III~C. 

\begin{figure}[b]
\centering 
\includegraphics[width=15cm]{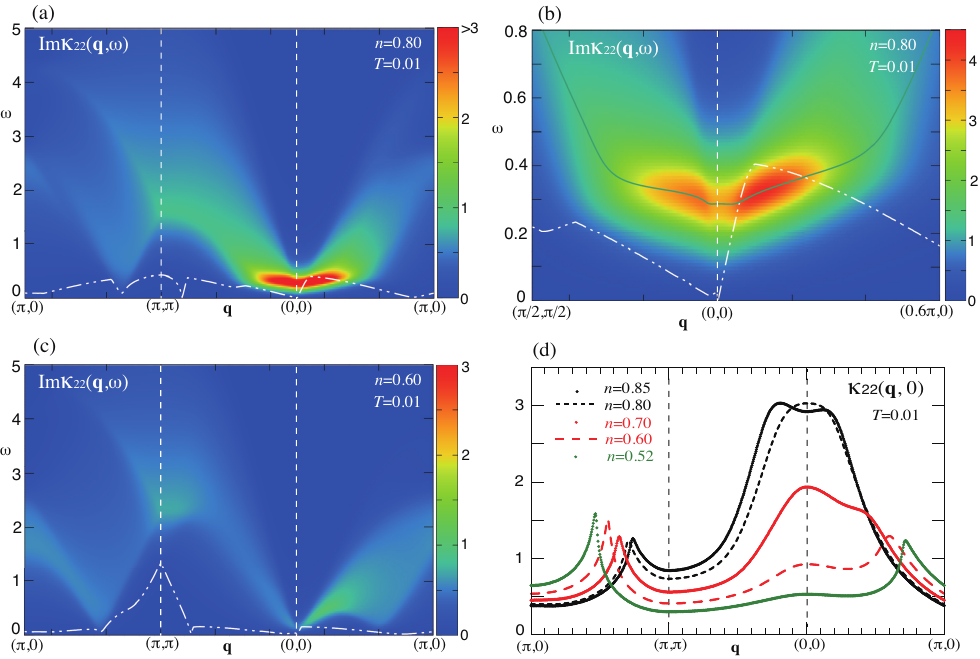}
\caption{$d$-wave bond-charge excitation spectra on the hole-doped side. (a) $\vq$-$\omega$ map of Im$\kappa_{22}(\vq, \omega)$ along the $(\pi,0)$-$(\pi,\pi)$-$(0,0)$-$(\pi,0)$ direction in the superconducting state ($T=0.01$) at $n=0.80$. The lower edge of the continuum is depicted by the dashed double-dotted curve. (b) Zoom of the low-energy region around $\vq=(0,0)$ in (a). The dispersion of the bond-charge resonance mode is shown by the green curve. (c) $\vq$-$\omega$ map of Im$\kappa_{22}(\vq, \omega)$ at $n=0.60$. (d) $\vq$ dependence of the static $d$-wave bond-charge susceptibility for several choices of the electron density in the superconducting state $(T=0.01)$.  
}
\label{kqw22-map-hole}
\end{figure}

Figure~\ref{kqw11-map-hole} is a $\vq$-$\omega$ map of Im$\kappa_{11}(\vq, \omega)$---the $s$-wave bond-charge excitations. In contrast to the electron-doped side [\fig{kqw-mapT005}(a)], Im$\kappa_{11}(\vq, \omega)$ exhibits incommensurate peaks at $\vq \approx (\pi, 0.7\pi)$ and $(0.8\pi, 0.8\pi)$ in low energy. The two incommensurate structures merge at $\vq=(\pi, \pi)$ and $\omega \approx 1.5$. Consequently a strong signal at $\vq=(\pi, \pi)$ is realized much higher energy than that on the electron-doped side  [\fig{kqw-mapT005}(a)]. Interestingly, a weak incommensurate structure appears again above $\omega \approx 3$. The spectrum along $(0,0)$-$(\pi, 0)$ exhibits qualitatively the same as that on the electron-doped side. Moreover, the spectrum does not change qualitatively with carrier doping [\fig{kqw11-map-hole}(b)], similar to the electron-doped side given in Figs.~\ref{kqw-mapT005}(a) and \ref{kqw-sc}(a).  Note that the continuum spectrum has a gap at $(\pi, \pi)$ even at $n=0.60$, which is larger than that at $n=0.75$, in spite of a smaller superconducting gap. This is due to the Fermi surface geometry as we already explained in the context of \fig{xqw-map-hole}.

\begin{figure}[b]
\centering 
\includegraphics[width=10cm]{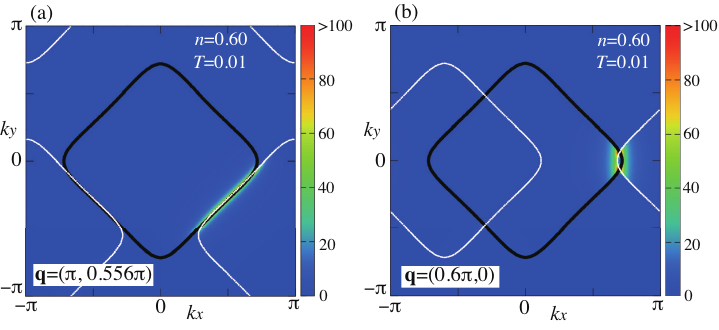}
\caption{Maps of \eq{integrant-kq} in the Brillouin zone for $\vq=(\pi, 0.556\pi)$ (a) and $(0.6\pi, 0)$ (b) at $n=0.60$; superconducting gap is set to zero for simplicity. The solid curve is the Fermi surface ($\xi_{\vk}=0$) and the white curve is the Fermi surface shifted by $-\vq$ and fulfills $\xi_{\vk+\vq}=0$.} 
\label{kq22-process-hole}
\end{figure}

The $d$-wave bond-charge excitations on the hole-doped side are very different from those in the electron-doped side. As shown in \fig{kqw22-map-hole}(a), Im$\chi_{22}(\vq, \omega)$ has large spectral weight around $\vq=(0, 0)$, not $(\pi, \pi)$, in a low energy region. This region is magnified in \fig{kqw22-map-hole}(b), indicating a bond-charge resonance mode inside the gap of the particle-hole continuum around $\vq \approx (0.1\pi, 0)$.  

Given that the $d$-wave bond-charge order at $\vq=(0,0)$ is nothing less than the order parameter of the $d$-wave Pomeranchuk instability \cite{yamase00a,yamase00b,metzner00}---the so-called electronic nematic order, \fig{kqw22-map-hole}(b) indicates the presence of substantial electronic nematic fluctuations on the hole-doped side.  This resonance mode is generated as a consequence of the competition of the nematic correlations and superconductivity as shown in Ref.~\cite{yamase04b}. 

For a higher doping [\fig{kqw22-map-hole}(c)], the charge resonance mode disappears, but still there is substantial low-energy spectral weight around $\vq=(0,0)$, forming a linear dispersion in the low-energy limit \cite{yamase04b}. A close inspection of \fig{kqw22-map-hole}(c) reveals additional low-energy structure at $\vq \approx (\pi, 0.6\pi)$ and $(0.6\pi,0)$. The former structure is recognized even in \fig{kqw22-map-hole}(a) although it is barely visible in the color scale there. 

Those features may be seen more clearly in the momentum dependence of the static susceptibility as shown in \fig{kqw22-map-hole}(d) for several choices of the electron density. The large peak at $\vq=(0,0)$ is an indication of the proximity to the electronic nematic instability especially for lower doping---the instability itself is hampered by the superconductivity for all electron density. While a peak is formed along the $(\pi, 0)$-$(\pi,\pi)$ direction for all electron density in \fig{kqw22-map-hole}(d), a peak around $\vq=(0.6\pi, 0)$ is realized only for high doping.

\begin{figure}[t]
\centering 
\includegraphics[width=7cm]{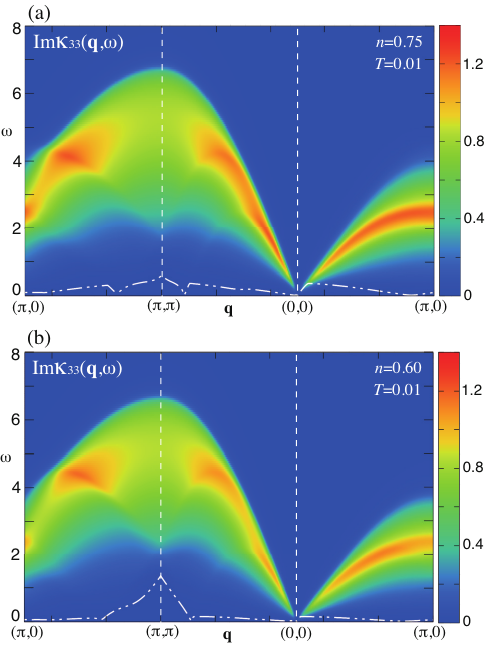}
\caption{$\vq$-$\omega$ maps of Im$\kappa_{33}(\vq, \omega)$ on the hole-doped side along the $(\pi,0)$-$(\pi,\pi)$-$(0,0)$-$(\pi,0)$ direction in the superconducting state ($T=0.01$): (a) $n=0.75$ and (b) $n=0.60$. The lower edge of the continuum is depicted by the dashed double-dotted curve.}
\label{kqw33-map-hole}
\end{figure}

To understand the origin of those peaks away from $\vq=(0,0)$, we consider the integrant of $\kappa_{0\, 22}(\vq, 0)$, which is given in the normal state by [see \eq{kqw1122}] 
\be
\gamma_{\vk_{+}}^{2} \frac{\tanh(\frac{\xi_{\vk+\vq}}{2 T}) -  \tanh(\frac{\xi_{\vk}}{2 T})}{\xi_{\vk+\vq} - \xi_{\vk}}
\label{integrant-kq}
\ee
after shifting $\vk$ by $\vq/2$; $\gamma_{\vk_{+}}=d_{\vk_{+}}$. Figure~\ref{kq22-process-hole} is a map of \eq{integrant-kq} in the Brillouin zone for a given $\vq$, showing that the states near the Fermi surface are responsible for the peak structure along the $(\pi,0)$-$(\pi, \pi)$ and $(0,0)$-$(\pi, 0)$ directions. Because the $d$-wave gap has nodes along the $k_{y}=\pm k_{x}$ axis, the peak near $\vq \approx (\pi, 0.6\pi)$ can survive even in the presence of the large superconducting gap, explaining a robust feature of the peak independent of the electron density [\fig{kqw22-map-hole}(d)]. For the peak near $\vq \approx (0.6\pi, 0)$ [\fig{kq22-process-hole}(b)], on the other hand, scattering processes between the edges of the electron-like Fermi surface is important and their low-energy processes are suppressed by the development of the $d$-wave superconducting gap there. Therefore, the peak near $\vq \approx (0.6\pi, 0)$ forms in a heavily doped region where the Fermi surface becomes electron-like and at the same time the superconducting gap becomes small. Considering how the Fermi surface shape evolves with carrier doping (inset in \fig{phase-hole}), we can explain the doping dependence of these peak positions seen in \fig{kqw22-map-hole}(d).

$\vq$-$\omega$ maps of Im$\kappa_{33}(\vq, \omega)$ are shown in \fig{kqw33-map-hole}. They are very similar to the result on the electron-doped side given in \fig{kqw-sc}(c). Furthermore, its doping dependence is very weak, similar to the electron-doped side.

\begin{figure}[bht]
\centering 
\includegraphics[width=15cm]{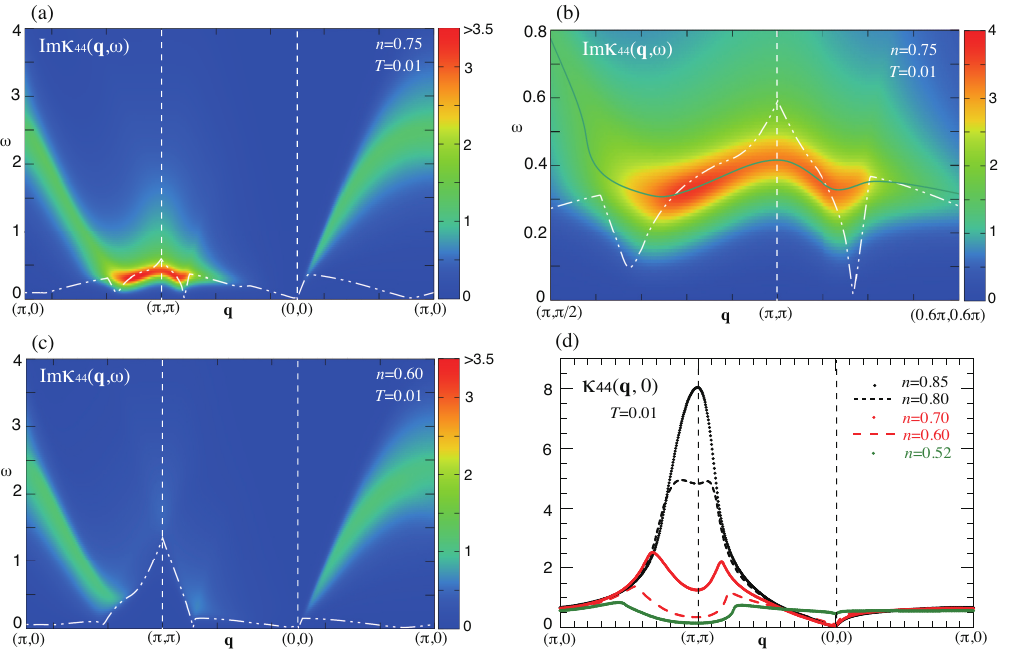}
\caption{Bond-charge excitation spectra with the form factor $\gamma_{\vk}=p^{-}_{\vk}$ on the hole-doped side. (a) $\vq$-$\omega$ map of Im$\kappa_{44}(\vq, \omega)$ along the $(\pi,0)$-$(\pi,\pi)$-$(0,0)$-$(\pi,0)$ direction in the superconducting state ($T=0.01$) at $n=0.75$. The lower edge of the continuum is depicted by the dashed double-dotted curve. (b) Zoom of the low-energy region around $\vq=(\pi, \pi)$ in (a). The dispersion of the bond-charge resonance mode is shown by the  green curve. (c) $\vq$-$\omega$ map of Im$\kappa_{44}(\vq, \omega)$ at $n=0.60$. (d) $\vq$ dependence of the static  susceptibility for several choices of the electron density in the superconducting state $(T=0.01)$.  }
\label{kqw44-map-hole}
\end{figure}

On the other hand, Im$\kappa_{44}(\vq, \omega)$ exhibits features qualitatively different from those on the electron-doped side [\fig{kqw-sc}(d)]. As shown in \fig{kqw44-map-hole}(a) it exhibits very strong weight around $\vq=(\pi, \pi)$ and forms a bond-charge resonance mode---\fig{kqw44-map-hole}(b) is a zoom of the resonance mode.  It is rather similar to the spin resonance mode given in \fig{xqw-map-hole}(b). The resonance mode implies that the proximity to the bond-charge instability with the form factor $\gamma_{\vk}=p_{\vk}^{-}$, namely the so-called $d$-density wave  \cite{chakravarty01} and flux phase \cite{cappelluti99}. Although strong spectral weight is realized around $\vq=(\pi, \pi)$ also on the electron-doped side [\fig{kqw-sc}(d)], that in \fig{kqw-sc}(d) is not related to a resonance mode. Upon carrier doping, the resonance mode is weakened and disappears. The spectrum leaves only an incommensurate structure especially along the direction $(\pi, \pi)$-$(\pi, 0)$ as shown in \fig{kqw44-map-hole}(c). The dispersion along the direction $(0,0)$-$(\pi, 0)$ seen in Figs~\ref{kqw44-map-hole}(a) and (c) is a peak of the particle-hole continuum, similar to that see in \fig{kqw-sc}(d). 

The incommensurate structure around $\vq=(\pi, \pi)$ and its doping dependence can be seen more clearly by studying the $\vq$ dependence of the static susceptibility as shown in \fig{kqw44-map-hole}(d)---the incommensurability is increased with doping. There is also an incommensurate peak along $(\pi, \pi)$-$(0,0)$ direction, although it is not clear in \fig{kqw44-map-hole}(c).

\section{Discussions}
We discuss results obtained on the electron-doped side in the first four subsections by comparing them with  experimental and theoretical results, showing overall good agreements. We next consider results obtained on the hole-doped side---they can capture several important features observed in experiments, but at the same time they also show some differences. As common features in both electron-doped and hole-doped cuprates, we then discuss the interpretation of the dispersion of magnetic excitations around $\vq=(0,0)$,  a role of usual charge excitations, and correlation effects beyond the present work.

\subsection{Magnetic excitations in electron-doped cuprates}
As we have shown in Figs.~\ref{xqw-mapT005} and \ref{xqw-sc}, the temperature dependence of the magnetic excitation spectra is very weak even when entering the superconducting state. This weak temperature dependence was actually reported in neutron scattering measurements of low-energy magnetic excitations for the optimally doped ${\rm Pr_{0.88}LaCe_{0.12}CuO_{4-\delta}}$ \cite{wilson06}.

\begin{figure}[bht]
\centering 
\includegraphics[width=6cm]{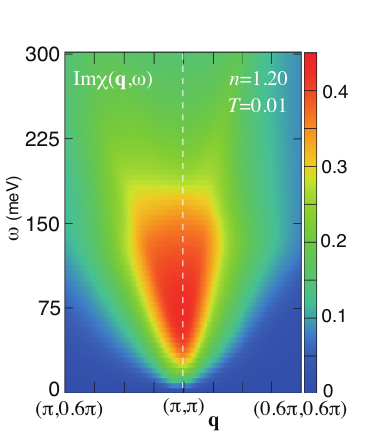}
\caption{Zoom of \fig{xqw-sc}(a) around $\vq=(\pi, \pi)$. The $\omega$ axis is given in the unit of meV by assuming $V_{m}=150$ meV \cite{bourges97b}.} 
\label{xqw-pen}
\end{figure}

The magnetic excitation spectrum shows pronounced spectral weight at $\vq=(\pi, \pi)$ in the low energy, especially close to the antiferromagnetic phase (\fig{xqw-af}). Even going away from the vicinity of the magnetic phase, the strong spectral weight is still retained around $\vq=(\pi, \pi)$ [\fig{xqw-sc}(a)]. A close look at the spectral map reveals that the spectral weight extends to higher energy ($\omega \sim 1$) and forms a pencil-tip structure as shown in \fig{xqw-pen}. This result captures nearly quantitatively the neutron scattering data obtained in $\text{Pr}_{0.89}\text{LaCe}_{0.11} \text{CuO}_{4}$ \cite{fujita12} and $\text{Pr}_{0.88}\text{LaCe}_{0.12} \text{CuO}_{4}$ \cite{wilson06b}. A careful comparison might suggest a slightly different feature close to zero energy---the pencil tip extends down to zero energy in Ref.~\cite{fujita12} whereas it appears with a ``gap'' around 25 meV in \fig{xqw-pen}. However, there is no gap in \fig{xqw-pen} and the intensity simply becomes weaker in the lower energy region.

It is remarkable to capture the experimental data in the electron-doped cuprates even with the present simple model. A key reason may lie in the shape of the Fermi surface. As we explain in \fig{xqw-process}, the excitation spectrum can be understood by considering scattering processes of electrons near the Fermi surface even in the high energy region. Hence, the band structure near the Fermi energy is crucial to the structure of the magnetic excitation spectrum. In the present study, we have chosen $t'$ and $t''$ to reproduce the Fermi surface observed in the electron-doped cuprates \cite{armitage10}. This can be a major reason why we have obtained a good agreement between the present theory and experiments in \fig{xqw-pen}. While Ref.~\cite{krueger07} claimed that the magnetic excitations in electron-doped cuprates cannot be understood in the fermiology approach, this contradicting conclusion could arise from the difference of the Fermi surface shape employed in their calculations. In fact, our results are broadly consistent with other theoretical works \cite{li03,onufrieva04}. Some theoretical studies \cite{yuan05,chen10,zhang13} reported the importance of the magnetic order to realize the commensurate magnetic fluctuations. Following those works, we would expect that our obtained magnetic excitation spectra do not change much even inside the magnetic phase.

In contrast to the hole-doped side [\fig{xqw-map-hole}(b)], the magnetic resonance mode is not developed. This is simply because the $d$-wave superconducting gap is small and furthermore electronic states near $\vk \sim (\pi/2, \pi/2)$, where $d$-wave gap nodes are present, are responsible to the peak at $\vq=(\pi, \pi)$ [see \fig{xqw-process}(b)], leading to a small gap of the particle-hole continuum on the electron-doped side. Although the ``resonance'' mode was reported in the electron-doped cuprates \cite{wilson06a,yu10}, the strongest signal is realized at $\vq=(\pi, \pi)$ inside the continuum \cite{zhao07}---it is not a true resonance mode and the present theory supports such a data. The experimental literature \cite{wilson06a,zhao07,yu10} consistently reported that the hourglass-like dispersive mode is not realized in the electron-doped cuprates, which is consistent with the present work.

\subsection{Bond-charge excitations in electron-doped cuprates} 
As shown in Figs.~\ref{kqw-mapT005}(b) and \ref{kqw-sc}(b), it is a characteristic feature  of the $d$-wave bond-charge excitations that exhibits softening along the $(0,0)$-$(\pi,0)$ direction. This feature was actually reported in the resonant x-ray scattering \cite{da-silva-neto15,da-silva-neto16} and RIXS \cite{da-silva-neto18} as a charge ordering tendency in electron-doped cuprates. Given that bond-charge excitations with the other symmetries do not have softening along the $(0,0)$-$(\pi, 0)$ direction (see \fig{kqw-mapT005}), it is reasonable to consider that the experimental data come from the $d$-wave bond-charge excitations. In fact, in-depth theoretical studies \cite{yamase15b,li17,yamase19} support such an interpretation. 

The present theory predicts that the appreciable spectral weight is present in the $d$-wave channel along the direction $\vq=(0,0)$-$(\pi, \pi)$, which also exhibits softening near $\vq=(0.7\pi, 0.7\pi)$ with decreasing temperature and doping as seen in Figs.~\ref{kqw-sc}(b), \ref{kq-T}(b) and \ref{kq-n}(b). 

Interestingly, Im$\kappa_{11}(\vq, \omega)$, Im$\kappa_{22}(\vq, \omega)$, and Im$\kappa_{44}(\vq, \omega)$ have strong spectral weight around $\vq=(\pi, \pi)$ (see Figs.~\ref{kqw-mapT005} and \ref{kqw-sc}). This can be a generic feature of bond-charge excitations. Since there is no experimental study of charge excitations around ${\bf q}=(\pi,\pi)$, it is worthwhile testing it in experiments.

\subsection{Bond-charge excitations: comparison with large-$N$ theory} 
\begin{figure}[ht]
\centering 
\includegraphics[width=12cm]{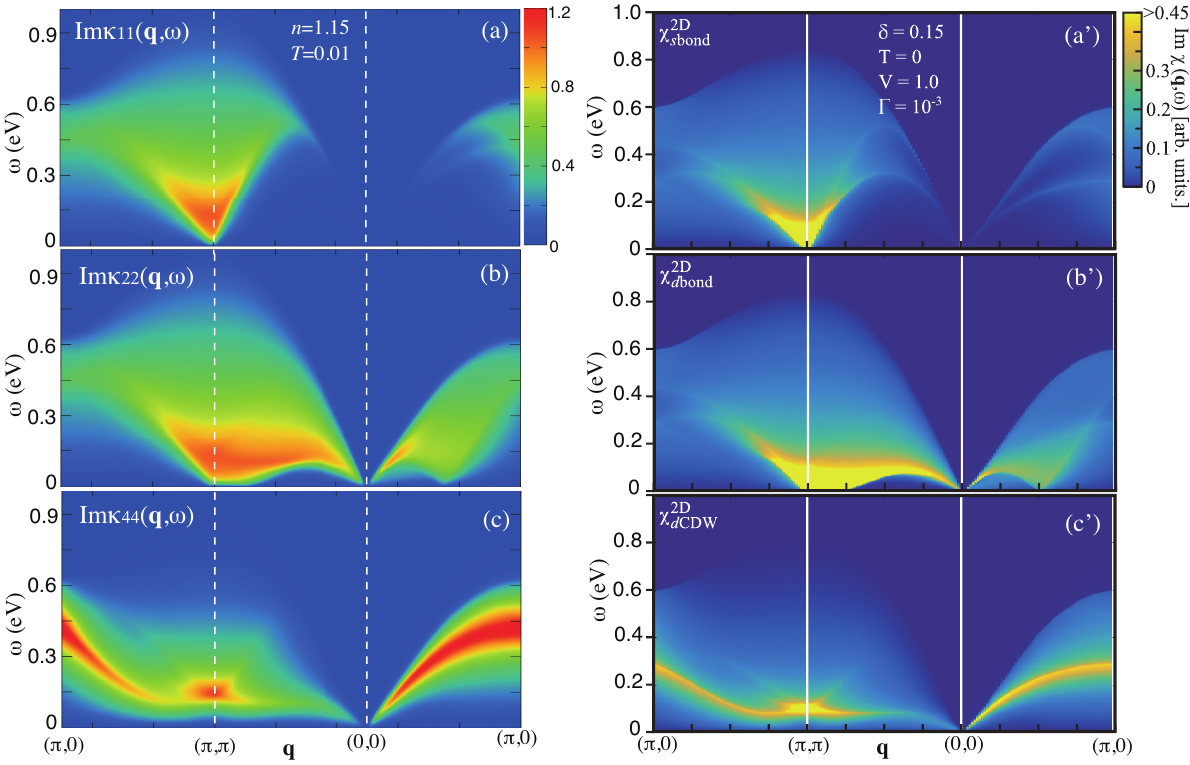}
\caption{Comparison of bond-charge excitation spectra obtained in the present theory (a)-(c) with those in a large-$N$ theory of the $t$-$J$ model (a')-(c') \cite{bejas17} at the same doping rate in the normal phase; (a) Im$\kappa_{11}(\vq, \omega)$, (b) Im$\kappa_{22}(\vq, \omega)$, and   (c) Im$\kappa_{44}(\vq, \omega)$ correspond to (a') $\chi^{\rm 2D}_{s{\rm bond}}$,  (b') $\chi^{\rm 2D}_{d{\rm bond}}$, and (c') $\chi^{\rm 2D}_{d{\rm CDW}}$, respectively. 
The $\omega$ axis is given in the unit of eV by assuming  $V_{m}=150$ meV in (a)-(c). In Ref.~\cite{bejas17}, $\delta$ is the electron doping rate, $V$ describes the nearest-neighbor Coulomb strength and $\Gamma$ is a broadening parameter; (a')-(c') are presented by modifying the original figures in Ref.~\cite{bejas17}. }
\label{largeN}
\end{figure}

Bond-charge excitations were studied in a large-$N$ theory of the $t$-$J$ model, especially for parameters appropriate for electron-doped cuprates \cite{bejas17}. Figure~\ref{largeN} is a comparison between our obtained spectra in the normal phase and  those obtained in the large-$N$ theory of the $t$-$J$ model at the same doping rate. Although the energy scale is slightly higher in our results, the overall features are strikingly similar to each other in spite of a fact that the effect of strong correlations is not included in our model whereas it is considered to some extent in the large-$N$ theory---hence the theoretical scheme is very different from each other.  This comparison suggests that the bond-charge excitation spectrum is captured semiquantitatively even in a weakly correlated model like our model and the band structure near the Fermi energy plays a crucial role to understand the structure of the spectrum as we have clarified in Figs.~\ref{kqw11-process}, \ref{kqw22-process}, \ref{kqw33-process}, and \ref{kqw44-process}.

\subsection{Comparison between magnetic and bond-charge excitations}
Both magnetic and bond-charge excitation spectra have similar energy scale as shown in Figs.~\ref{xqw-mapT005} and \ref{kqw-mapT005}. This is because motivated by the theoretical observation that the bond-charge interaction originates from the spin-spin interaction \cite{fczhang88a, affleck88,marston89,morse91,cappelluti99,yamase00a,yamase00b,lee06,bejas12,sachdev13} and also by a similarity to the slave-boson theory of the $t$-$J$ model \cite{lee06}  [see also the paragraph above \eq{MF-chi}], we have chosen the interaction strength to be controlled by the magnetic interaction $V_{m}$. 

The magnetic excitation spectra (\fig{xqw-mapT005}) is similar to that of the $s$-wave bond-charge excitations [\fig{kqw-mapT005}(a)] in that the strong spectral weight is concentrated around $\vq=(\pi, \pi)$ in a low-energy region. While the dispersive feature is observed around $\vq=(0,0)$ in the magnetic excitations, a  corresponding feature is not clearly recognized in the $s$-wave bond-charge excitations. 

An overall comparison between Figs.~\ref{xqw-mapT005} and \ref{kqw-mapT005} shows that the spectral weight of the bond-charge excitations is stronger for any symmetry and extends in a much wider $\vq$-$\omega$ space than the magnetic excitations. This feature is mostly prominent in the $d$-wave bond-charge excitations shown in \fig{kqw-mapT005}(b). Given that the magnetic excitation spectrum was successfully detected in electron-doped cuprates \cite{wilson06,wilson06a,zhao07,fujita12}, there seems a good chance to detect by RIXS also the widely spreading bond-charge excitations, especially in the $d$-wave channel. 

Since the characteristic momentum of the magnetic excitation spectrum is $\vq=(\pi, \pi)$, the magnetic fluctuations are not expected to couple to some charge fluctuations with a finite momentum. Rather typical momenta of both magnetic and bond-charge excitations are determined independently by the Fermi surface geometry as seen in Figs.~\ref{xqw-process}(b)-(d), \ref{kqw11-process}(b), \ref{kqw22-process}(b)-(e), \ref{kqw33-process}(b), and \ref{kqw44-process}(b),(c). A possible coupling of magnetic fluctuations and charge fluctuations discussed in Ref.~\cite{da-silva-neto18}, if any, could be an effect beyond the present theory.  

\subsection{Similarities and differences to hole-doped cuprates} 
The present effective model does not consider the strong correlation effect originating from the local constraint that double occupancy of electrons is prohibited at any lattice site. Nonetheless, the model captures the phase diagram, the magnetic excitations, and a softening of bond-charge spectrum around $\vq=(0.5\pi,0)$ observed in electron-doped cuprates. Furthermore, the bond-charge excitation spectra computed in the present model yield results very similar to those in the large-$N$ theory of the $t$-$J$ model, where the strong correlation effect is incorporated. Therefore it is tempting to clarify to what extent the present model can capture the hole-doped cuprates. 

The phase diagram obtained in the present theory (\fig{phase-hole}) captures the qualitative feature of the phase diagram in hole-doped cuprates in the sense that $d$-wave superconductivity close to the antiferromagnetic phase \cite{lee06,armitage10,scalapino12,keimer15}. However, there are important differences. First the onset temperature of the antiferromagnetism is comparable to that of the superconductivity, whereas in reality the former is much higher than the latter. Second, both antiferromagnetic and superconducting phases extend much higher doping than the reality. Third, the superconducting phase does not have a dome-shaped form. While we can tune the interaction strength as free parameters in the effective model [\eq{Hamiltonian}], we did not find a parameter set that improves substantially the agreement with the experimental phase diagram. Furthermore, the so-called pseudogap is not captured in the present model. 

The hole pockets in the antiferromagnetic phase are observed clearly in multilayered cuprates \cite{kunisada20,kurokawa23}. In the paramagnetic phase, we obtain the hole-like Fermi surface in $n \gtrsim 0.8$ and the Fermi surface topology changes to be electron-like upon further hole doping. Such a change of the topology is reported at $n\approx 0.78$ in La-based cuprates \cite{yoshida06}.

The hourglass dispersion of the magnetic excitation spectrum including the spin resonance mode [\fig{xqw-map-hole}(b)] well captures the essential feature of experimental data \cite{fujita12}. In particular, when we choose the realistic value of the magnetic interaction strength $V_{m}=100 \sim 150$ meV \cite{bourges97b}, the absolute energy is also comparable with the experimental data. Note, however, the doping rate itself is too high in the present theory, which is because our  phase diagram (\fig{phase-hole}) does not reproduce quantitative features of the experimental phase diagram \cite{keimer15,scalapino12,lee06,armitage10}. In contrast to the electron-doped side, the incommensurate magnetic structure is usually realized on the hole-doped side (\fig{xqw-map-hole}), which is consistent with a generic feature observed in the hole-doped cuprates \cite{birgeneau06,fujita12}. These results imply the importance of fermiology to understand the magnetic excitations in hole-doped cuprates and confirm various theoretical studies \cite{bulut96,norman00,manske01,schnyder04,jia14, tanamoto94,brinckmann99,yamase01,brinckmann02,onufrieva02,li02,sega03,yamase06,yamase07,li16, si93,zha93,liu95,millis96,kao00,chubukov01,james12}---we do not find essentially new insights that should be added  except for the spectrum around $\vq=(0,0)$ discussed in the next subsection.

In contrast to the case of the magnetic excitations, our results of bond-charge excitations cannot capture charge excitations in hole-doped cuprates. In experiments, the charge ordering tendency is reported around the direction $(0,0)$-$(\pi, 0)$ \cite{ghiringhelli12,chang12,achkar12,blackburn13,blanco-canosa14,comin14,tabis14,da-silva-neto14,hashimoto14,peng16,chaix17,arpaia19,yu20,wslee21,lu22,arpaia23}. In the present theory it is only the $d$-wave bond-charge excitations that show a similar feature, but at rather high-doping rate [Figs.~\ref{kqw22-map-hole}(c) and (d)]. The origin of this softening comes from scattering between the edge of the electron-like Fermi surface [\fig{kq22-process-hole}(b)] and thus its wavevector becomes larger with further carrier doping  [\fig{kqw22-map-hole}(d)].  In the hole-doped cuprates, however, the charge-ordering tendencies were reported around $n=0.88$ \cite{ghiringhelli12,chang12,achkar12,blackburn13,blanco-canosa14,comin14,tabis14,da-silva-neto14,hashimoto14,peng16,chaix17,arpaia19,yu20,wslee21,lu22,arpaia23}, its wavevector becomes smaller with carrier doping, and the Fermi surface is expected hole-like in that  doping region. Hence the softening of the $d$-wave bond-charge excitations in \fig{kqw22-map-hole}(c) cannot explain the experimental data qualitatively. Recalling that the present theory is formulated in a weak coupling framework, strong correlation effects including the pseudogap seem crucial to understand the charge excitations observed in the hole-doped cuprates \cite{loret19}. Note that the charge excitations observed around $\vq=(0,0)$ in hole-doped cuprates are understood in terms of plasmons, namely usual charge excitations, not bond-charge ones \cite{greco16,greco19,greco20,nag20,hepting22,hepting23}. 

Bond-charge excitations on the hole-doped side predict the bond-charge resonance modes near $\vq=(0,0)$ in the $d$-wave channel [\fig{kqw22-map-hole}(b)] and $\vq=(\pi,\pi)$ in the $p_{\vk}^{-}$ channel [\fig{kqw44-map-hole}(d)] in the $d$-wave superconducting state. At the moment, there is no indication of the charge resonance mode in the hole-doped cuprates.

\subsection{Interpretation of paramagnon dispersion in cuprates}
The present theory provides an important caveat about the interpretation of the magnetic excitations. We have found the sharp dispersive feature of magnetic excitations around the momentum $\vq=(0,0)$ in both electron-doped (\fig{xqw-sw}) and hole-doped sides (\fig{xqw-map-hole}). This dispersion beers a striking resemblance to that found in both electron-doped \cite{ishii14,wslee14} and hole-doped cuprates \cite{braicovich10,letacon11,letacon13,dean13,wakimoto15,monney16,ivashko17,meyers17,chaix18,peng18a,robarts19,wang22} and can be analyzed successfully in terms of the linear spin-wave theory as we have presented in \fig{xqw-sw}. Hence as seen in many experimental literatures \cite{braicovich10,letacon11,letacon13,dean13,ishii14,wslee14,wakimoto15,monney16,ivashko17,meyers17,chaix18,peng18a,robarts19,wang22}, it might seem reasonable to interpret it as {\it magnon}-like excitations in the {\it paramagnetic}  phase, namely {\it paramagnons}. However, our obtained dispersion can be interpreted more appropriately as a peak structure of individual particle-hole excitations with short-range spin correlations. Some collective features associated with the magnetic ordering are seen only around $\vq=(\pi,\pi)$ in the present theory (\fig{xqw-sw}) and the dispersive feature around $\vq=(0,0)$ does not change much upon approaching the magnetic phase, which is clear by making a comparison between Figs.~\ref{xqw-mapT005} and \ref{xqw-af}(a). This feature is in line with RIXS data \cite{dean13,dean13a,chaix18} as well as calculations in the Hubbard model \cite{jia14}. Moreover, as seen in \fig{xq}, the static susceptibility does not show a structure around $\vq=(0,0)$.

Our interpretation is shared with some experimental \cite{minola17,monney16} and theoretical studies \cite{zeyher13,guarise14,benjamin14,kanasz16,zhang22}.  Although a theoretical prediction given in Refs.~\cite{benjamin14,kanasz16}---fluorescencelike shift of the RIXS spectrum with changing the photon energy---was not confirmed by experiments especially in the underdoped region \cite{minola15}, the dispersion of the high-energy magnetic excitations is realized close to the upper boundary of the continuum spectrum in the present theory (Figs.~\ref{xqw-mapT005} and \ref{xqw-sw}). Hence, our dispersion would show a weak fluorescencelike shift if any, in favor of the experimental data \cite{minola15}. Moreover, a recent theoretical analysis \cite{zhang22} pointed out an importance of short-range spin correlations to understand the experimental data, in agreement with the present theory.

In Figs.~\ref{xqw-mapT005}, \ref{xqw-sw}(a), and \ref{xqw-map-hole}(a)(d), the dispersion along the $(0,0)$-$(\pi, 0)$ direction is much clearer than that along the $(0,0)$-$(\pi, \pi)$ direction. This anisotropy of intensity \cite{monney16} as well as anisotropy damping \cite{peng18a,robarts19} was reported in RIXS, but not in Ref.~\cite{chaix18}.

\subsection{Role of usual charge excitations} 
The vertex of the bond-charge interaction has a momentum dependence [\fig{diagram1}(a)]. In this sense, it is sharply  different from the usual charge density interaction, which has a momentum-independent vertex. Nonetheless, the usual charge fluctuations are also expected to be present in actual materials. Hence, although it is beyond the scope of the present model [\eq{Hamiltonian}], it should be valuable to comment on a possible role of the usual charge excitations by referring to literature. 

Both usual charge fluctuations and bond-charge fluctuations were analyzed in a large-$N$ theory of the $t$-$J$ model by including the long-range Coulomb interaction as well as a realistic layer structure of cuprate superconductors. Reference \cite{bejas17} revealed a dual structure of the excitation spectrum in energy space---the bond-charge excitations appear in the low-energy region scaled by $J$ inside the particle-hole continuum whereas the usual charge excitations are realized as plasmons above the continuum. It should be noted that the usual charge excitations are also  present in the low-energy region as particle-hole excitations, but their spectral weight is weak and the bond-charge excitations are dominant there. In addition, there is a coupling between the usual charge fluctuations and the bond-charge fluctuations. However, explicit calculations in the $t$-$J$ model found that such a coupling is weak \cite{bejas17}. 

\subsection{Correlation effects beyond the RPA} 
We have employed a weak coupling model [\eq{Hamiltonian}] to analyze both magnetic and bond-charge excitations on an equal footing. While reasonably good agreements are found especially for electron-doped cuprate superconductors, this does not imply an importance of weak correlation effects, but rather an importance of the band structure near the Fermi energy even at relatively high energy; see Figs.~\ref{xqw-process}, \ref{kqw11-process}, \ref{kqw22-process}, \ref{kqw33-process}, and \ref{kqw44-process}. A possible reason why the agreement is better on the electron-doped side may be related to a fact that the pseudogap is much less pronounced in the electron-doped cuprates \cite{armitage10} and thus the band structure near the Fermi energy is captured appropriately even in the present theoretical scheme. Since the cuprate superconductors are doped Mott insulators, electron correlation effects beyond the RPA should be considered carefully for both magnetic and bond-charge excitations for further in-depth studies. Interesting subjects would be, to name a few,  i) how vertex corrections modify the effective interactions in the bond-charge and spin channels---a possible condition to realize a bond-charge instability, ii) how the self-energy modifies the one-particle excitation spectrum—spectral weight distribution along the Fermi surface and a possibility to form a pseudogap, and iii) mutual interplay of bond-charge and magnetic fluctuations.

\section{Conclusions}
We have studied an effective model describing magnetic, bond-charge, and pairing interactions on a square lattice by choosing model parameters appropriate to electron-doped cuprate superconductors. The obtained phase diagram shows the antiferromagnetic phase close to the half-filling and the $d$-wave superconducting phase nearby---a bond-charge ordered phase is not realized (\fig{phase}). This phase diagram reproduces essential features of the phase diagram in electron-doped cuprates. Magnetic excitations have strong low-energy spectral weight around $(\pi, \pi)$ and the spectral weight forms a pencil-tip shape (Fig.~\ref{xqw-pen}). In addition, their temperature dependence is very weak (Figs.~\ref{xqw-mapT005} and \ref{xqw-sc}).  These features agree with neutron scattering data. Furthermore, the high-energy spin excitations are realized around $(0,0)$ and are reminiscent of a spin-wave dispersion although the system is in the paramagnetic phase (\fig{xqw-sw}). RIXS experiments detected a dispersive feature very similar to ours and called it paramagnons. However, our dispersion should not be interpreted as paramagnons because the essential feature around $(0, 0)$ is already captured by Im$\chi_{0}(\vq, \omega)$, namely without the magnetic interaction [\fig{xqw-sw}(b)]. The $d$-wave bond-charge excitations show softening around $(\pi/2, 0)$  [\fig{kqw-mapT005}(b)].   This feature is actually observed in resonant x-ray scattering studies. Both $s$- and $d$-wave bond-charge excitations have large spectral weight around $(\pi, \pi)$ in low energy and spread widely from $(\pi, \pi)$ [Figs.~\ref{kqw-mapT005}(a) and (b)]. The excitation spectra do not change much even if the superconducting instability occurs at low temperatures in both spin (\fig{xqw-sc}) and bond-charge channels (\fig{kqw-sc}). The present model has an advantage to handle both magnetic and bond-charge excitations on an equal footing in the same approximation scheme. We find that the spectral weight of bond-charge excitations is typically much larger than that of magnetic excitations, except for the very low-energy region around $(\pi,\pi)$ when the system is close to the magnetic instability. 

While our model is very simple, it can capture not only the phase diagram but also magnetic and bond-charge excitations in electron-doped cuprates very well. A possible reason may lie in the shape of the Fermi surface, which we have chosen typical of that in electron-doped cuprates---we find that electrons near the Fermi surface are largely involved in determining the structure of magnetic and bond-charge excitations even in a high-energy region. Predictions  from the present theory are the presence of bond-charge excitations around $(\pi, \pi)$ (Figs.~\ref{kqw-mapT005} and \ref{kqw-sc}) and additional softening in the $d$-wave channel in the $(\pi, \pi)$-$(\pi/2, \pi/2)$ region at low temperatures near the magnetic phase [Figs.~\ref{kqw-sc}(b), \ref{kq-T}(b), and \ref{kq-n}(b)]. Those experimental tests should be important to establish the concept of bond-charge excitations in cuprates. 

The present calculations are extended to the hole-doped side. In contrast to the electron-doped side, the  incommensurate peak structure is realized in low energy around $(\pi, \pi)$ in both magnetic (\fig{xqw-map-hole}) and $s$-wave bond-charge excitations (\fig{kqw11-map-hole}). The spectrum of the $d$-wave bond-charge excitations change dramatically and exhibits a strong spectral weight around $(0,0)$, showing the electronic nematic tendency (\fig{kqw22-map-hole}). In the superconducting phase, the magnetic excitation spectrum exhibits the spin resonance mode with the so-called hourglass dispersion [\fig{xqw-map-hole}(b)], capturing the experimental data. The bond-charge resonance modes are also found in the $d$-wave channel around $(0,0)$ [\fig{kqw22-map-hole}(b)] and the $p^{-}_{\vk}$ channel around $(\pi, \pi)$ [\fig{kqw44-map-hole}(b)]. While the magnetic excitation spectra obtained in the present model capture neutron scattering data in the hole-doped cuprates, the spectra of bond-charge excitations do not capture the major feature observed in hole-doped cuprates, suggesting the importance of strong correlation effects including the pseudogap especially to understand the charge excitations. In fact, in spite of freedoms to choose independently the three different interaction values in our model [\eq{Hamiltonian}], the obtained phase diagram (\fig{phase-hole}) cannot capture the experimental phase diagram except for a feature that the $d$-wave superconducting phase is realized close to the antiferromagnetic phase. An interesting outcome on the hole-doped side is the presence of bond-charge resonance modes. In particular, it might be worthwhile to explore the $d$-wave channel in RIXS experiments, since the mode is expected if the system is close to the electronic nematic instability, which is actually implied in hole-doped cuprates by   various techniques: neutron scattering \cite{hinkov04,hinkov07,hinkov08,haug10}, electronic Raman scattering \cite{auvray19}, angle-resolved photoemission \cite{nakata21}, Compton scattering \cite{yamase21}, Nernst coefficient \cite{daou10,cyr-choiniere15}, and magnetic torque \cite{sato17}. 

A concept of the bond charge has been recognized in theory and it originates from the spin-spin interaction \cite{fczhang88a, affleck88,marston89,morse91,cappelluti99,yamase00a,yamase00b,lee06,bejas12,sachdev13}, leading to its dynamics with the energy scale comparable to the magnetic one. We hope that the present theoretical work will be a pedagogical guide to explore bond-charge excitations in metals near the antiferromagnetic phase such as cuprates and others.  


\acknowledgments
The authors thank M. Fujita,  A. Greco, M. Hepting, M. Minola, and D. Vilardi  for valuable discussions. H.Y. is indebted to warm hospitality of Max-Planck-Institute for Solid State Research and was supported by JSPS KAKENHI Grant No.~JP20H01856 and World Premier International Research Center Initiative (WPI), MEXT, Japan.


\appendix





\section{All components of the bond-charge excitation spectra}

\begin{figure}[tb]
\centering 
\includegraphics[width=16cm]{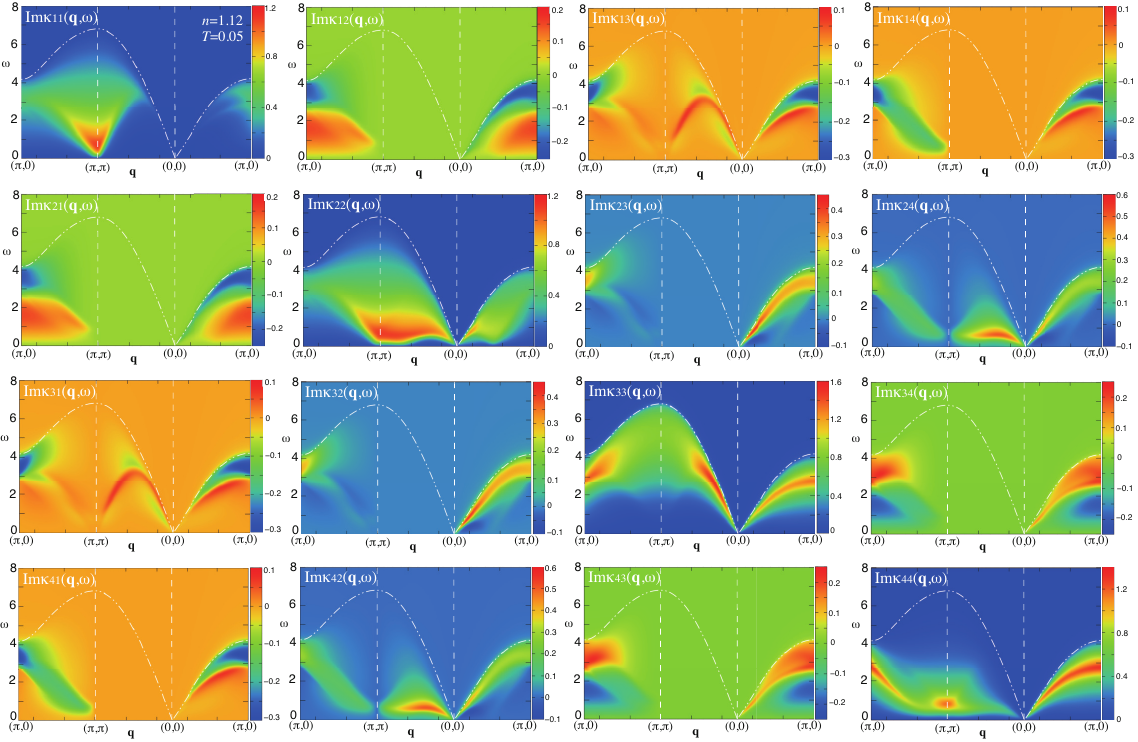}
\caption{$\vq$-$\omega$ maps of all components of the bond-charge excitation spectrum along the $(\pi,0)$-$(\pi,\pi)$-$(0,0)$-$(\pi,0)$ direction at $n=1.12$ in the normal state ($T=0.05$). The ``matrix'' is symmetric in the sense that Im$\kappa_{ji}$=Im$\kappa_{ij}$. The symmetry of the vertex $\hat{\Gamma}_{i}(\vk)$ yields Im$\kappa_{12}$=  Im$\kappa_{14}$= Im$\kappa_{23}$= Im$\kappa_{34}$=0 along the direction $(\pi, \pi)$-$(0,0)$. The off-diagonal components of Im$\kappa_{ij}$ can be negative---hence they are not physical on their own---whereas the diagonal ones are always positive as they should be. The dashed double-dotted curve is the upper boundary of the continuum spectrum.}
\label{kqw-mapall}
\end{figure}

While we have presented the diagonal parts of the bond-charge excitation spectra in the main text,  the full set of the spectra is shown in \fig{kqw-mapall}. First, the ``bond-charge matrix'' is symmetric, namely Im$\kappa_{ij}(\vq, \omega) =$Im$\kappa_{ji}(\vq, \omega)$. Second, while the spectral weight of the diagonal components is positive in $\omega>0$ as they should be, the off-diagonal components can become negative. In this sense, the off-diagonal components alone cannot be physical quantities. Third, because of the symmetry, the spectral weight along the direction $\vq=(\pi, \pi)$-$(0,0)$ vanishes in Im$\kappa_{12}$,  Im$\kappa_{14}$,  Im$\kappa_{23}$, and Im$\kappa_{34}$.  

\begin{figure}[ht]
\centering 
\includegraphics[width=16cm]{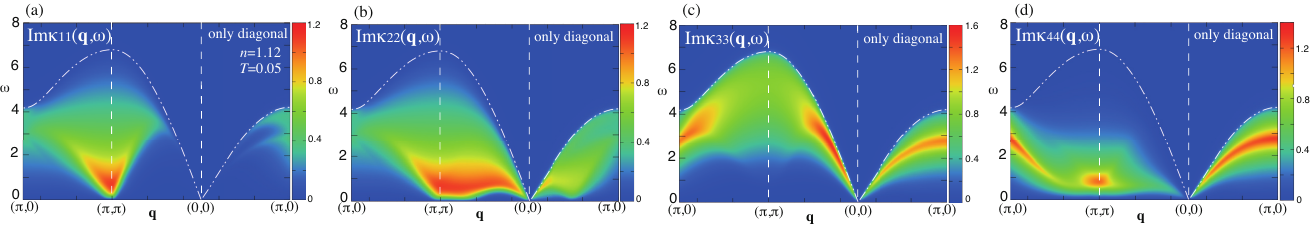}
\caption{$\vq$-$\omega$ maps of the bond-charge excitation spectrum without considering coupling to the other symmetries [see \eq{kappa-diag}]: (a) Im$\kappa_{11}(\vq, \omega)$, (b) Im$\kappa_{22}(\vq, \omega)$, (c) Im$\kappa_{33}(\vq, \omega)$, and (d) Im$\kappa_{44}(\vq, \omega)$ along the $(\pi,0)$-$(\pi,\pi)$-$(0,0)$-$(\pi,0)$ direction at $n=1.12$ in the normal state ($T=0.05$). These results should be compared with the diagonal components in \fig{kqw-mapall}, respectively, showing a weak coupling among different symmetries. The dashed double-dotted curve is the upper boundary of the continuum spectrum.}
\label{kqw-map-diag}
\end{figure}

An important question is how much the off-diagonal components have an impact on physical quantities. To check this, we compute the diagonal components by discarding all off-diagonal components by using the following expression with the same parameters: 
\be
\hat{\kappa}_{i i}(\vq, \omega)=(1+V_{b}\hat{\kappa}_{0 \, i i}(\vq, \omega))^{-1} \hat{\kappa}_{0\, i i}(\vq, \omega)\,.
\label{kappa-diag}
\ee
Obtained results are presented in Figs.~\ref{kqw-map-diag} (a)-(d), which should be compared with the diagonal results given in \fig{kqw-mapall}---those are almost the same as each other. This demonstrates that the effect of a coupling to  bond-charge fluctuations among different symmetries are weak. 

\bibliographystyle{apsrev4-1}
\bibliography{main} 

\begin{thebibliography}{154}
\bibitem [{mis()}]{misc-LSCO}%
  \BibitemOpen
  \href@noop {} {}\bibinfo {note} {The charge order observed in La-based
  cuprates is often discussed in terms of the spin-charge stripe order
  \cite{tranquada95}. While a possible connection to the newly observed charge
  ordering tendency around $q_{x} \sim 0.6\pi$ in other hole-doped cuprates
  provides a very interesting subject, this issue is beyond the scope of the
  present work.}\BibitemShut {Stop}%
\end{thebibliography}

\end{document}